# Role of lattice structure and breaking of antiferromagnetic spin order in enhancement of ferromagnetic, electronic, and magneto-electric properties in $Fe_{2-x}Sc_xO_3$ system

[a]*R. N. Bhowmik*[*], [a]*Bipin Kumar Parida*, [b,c]*Amit Kumar*, [d]*P.D. Babu*, [b,c]*S. M. Yusuf*

[a]Department of Physics, Pondicherry University, R. V Nagar, Kalapet-605014, India.

[b]Solid State Physics Division, Bhabha Atomic Research Centre, Mumbai- 400-085, India

[c]Homi Bhabha National Institute, Anushaktinagar, Mumbai 400094, India

[d]UGC-DAE Consortium for Scientific Research, Mumbai Center, R-5 Shed, BARC, Trombay, Mumbai, 400085, India

[*] Corresponding author: rnbhowmik.phy@pondiuni.edu.in

Abstract

The strategy of breaking antiferromagnetic (AFM) ground state in $\alpha$-$Fe_2O_3$ by doping non-magnetic $Sc^{3+}$ ($3d^0$) ions at the $Fe^{3+}$ ($3d^5$) sites has been used in $Fe_{2-x}Sc_xO_3$ system (x = 0.2-1.0). The material has been stabilized in single-phase (rhombohedral $\alpha$-$Fe_2O_3$) or mix-phase (rhombohedral $\alpha$-$Fe_2O_3$ and cubic $Sc_2O_3$-types) structure by varying Sc content and heat treatment temperature. Neutron diffraction confirmed perturbed AFM ground state down to low temperature with magnetic moment ~ 2.75-4.68 μB/Fe site and Morin transition ~ 260 K. DC magnetic measurement showed magnetic coercivity 0.2 to 6 kOe. The material showed transformation from insulating state (conductivity $10^{-14}$-$10^{-10}$ S/cm and polarization 0.5-2 μC/cm$^2$) to high conductive state (conductivity ~ $10^{-10}$ -$10^{-7}$ S/cm and polarization > 2 μC/cm$^2$) above the Morin transition. The room temperature measurements showed maximum current density 35-186 μA/cm$^2$, electric polarization 2.7-15.6 μC/cm$^2$, magneto-electric voltage up to 5 mV with coupling constant $\alpha_{ME}$ ~ 0.62-10.11 mV/Oe.cm and magneto-conductance up to 90 %. The results will open the door for suitably modifying the lattice-structure, magnetic spin order, and charge-spin coupling in hematite based material and their application in low power spintronic devices.

## 1. Introduction

Hematite (α-$Fe_2O_3$) is a Rhombohedral structured oxide with non-centrosymmetric $R\bar{3}c$ phase, which showed collinear AFM spin order (along *c*-axis) at the ground state below Morin transition (260 K) and a weak FM for temperatures above Morin transition ($T_M$) due to off-plane spin canting [1]. The hematite is disadvantageous for practical applications for its low magnetic moment, electrically insulator and non-ferroelectric due to non-polar (centrosymmetric and non-ferroelectric) nature of $R\bar{3}c$ space group. The doping of non-magnetic elements Ga and Al have a tremendous effect on enhancing both the ferromagnetic moment and ferroelectric polarization [2-4]. The alloying of non-polar α-$Fe_2O_3$ and α-$Al_2O_3$ (both in R-3c space group) showed polar orthorhombic structure in $AlFeO_3$. This shows the possibility of polar phase (R3c group) formation in Rhombohedral structure from the co-ordering of two cations (A and B) at the M sites of centrosymmetric (R-3c) $M_2O_3$ compounds. Apart from the chemical nature of doping ions, material synthesis, metal-oxygen bonding between 3d-2p orbitals can play significant roles for forming polar or non-polar phase or structural instability at the atomic scale. It is theoretically predicted as well as experimentally observed that the high symmetry nonpolar (R-3c) phase can transform into low symmetry polar (R3c) phase in perovskite ($ABO_3$) -type oxides, irrespective of the electronically metals, semiconductors and insulators [5]. The $LiOsO_3$ is a metal oxide, where displacement of the Li atom from its high temperature non-polar ($R\bar{3}c$ and high resistive) phase caused a ferroelectric-polar (and high conductive) phase at temperature below 140 K [6, 7]. Recent reports [8, 9] indicated that a noncentrosymmetric crystal structure is not necessarily be the prerequisite for exhibiting piezoelectric and ferroelectric properties. The electric field–induced lattice distortion can generate substantially large piezoelectric effects in centrosymmetric oxides.

$ScFeO_3$ is a newly discovered multiferroic compound [10]. It showed *G*-type AFM/weak FM order of $Fe^{3+}$ ($3d^5$) spin moments with magnetic transition at 545 K and non-centrosymmetric polar (R3c) phase due to off-centering of $Sc^{3+}$ ($3d^0$) ions predicted ferroelectric polarization up to ~105 μC/cm$^2$. Hamasaki et al. [11] demonstrated that $ScFeO_3$ lattice is located at the cross point of various meta-stable (bixbyite-type and corundum-type) structures. The alloying of $Sc_2O_3$ and $Fe_2O_3$ during formation of $ScFeO_3$ can stabilize either in bixbyite-type (cubic) non-polar phase due to a random distribution of the Sc and Fe ions at the lattice sites [12] or in polar (*R3c)* phase due to proper order of the $Sc^{3+}$ and $Fe^{3+}$ ions from the instability of the non-polar $R\bar{3}c$ phase [13]. The polar phase of $ScFeO_3$ exhibited ferroelectric polarization ~5 μC/cm$^2$ [14]. Kawamoto et. al. [15] argued that orthorhombic phase (*Pnma* space group), instead of polar R-3c corundum phase, is stabilized in $ScFeO_3$ at high pressure (~ 15 GPa) and high temperature (800 °C). Kim et al. [13] showed that R3c (hexagonal or pseudo-cubic) phase at atmospheric pressure can be transformed into orthorhombic phase (space group *Pnma*) at pressure ~ 37.4 kbar.

The incorporation of non-magnetic Sc at the Fe sites of α-$Fe_2O_3$ structure has modified the magnetic and electrical properties, apart from the observation of $Sc_2O_3$-type secondary phase [16]. Some of the $Fe_{2-x}Sc_xO_3$ samples prepared by mechanical alloying of $Sc_2O_3$ and $Fe_2O_3$ and post-heat treat showed ferroelectric polarization -like loop at room temperature, although $R\bar{3}c$ phase of α-$Fe_2O_3$ or cubic bixbyite (space group $Ia\bar{3}$) phase of $Sc_2O_3$ [12, 17] is non-polar. Hence, a detailed study is essential to examine the roles of Sc atoms in breaking the AFM spin order and modifying the electronic structure and charge-spin coupling properties in AFM oxide system [18, 19]. In this work, we have extended the neutron diffraction, dc magnetization, current-voltage loop and electric polarization measurements from room temperature down to 5 K to understand a correlation between the lattice structure, lattice disorder (role of $Sc_2O_3$-type (non-magnetic) secondary phase), breaking of AFM ground state, electrical conductivity and polarization in $Sc_2O_3$−$Fe_2O_3$ alloyed system. Additionally, magnetic field induced voltage and electrical conductivity have been measured at room temperature to confirm charge-spin coupling effect at room temperature, which is important for use of antiferromagnetic oxides in low power consuming spintronic devices.

## 2. Experimental

### *2.1. Material preparation*

Details of the material preparation and structural phase characterization of the $Fe_{2-x}Sc_xO_3$ system (x = 0.2, 0.5 and 1) have been discussed in previous work, defined as second method (M2) of mechanical alloying [16]. In this work, we have prepared nearly 6 g of alloyed powder for each composition by mechanical alloying of a stoichiometric mixture of α-$Fe_2O_3$ and $Sc_2O_3$ powders. The mechanical alloying was carried out by employing FRITSCH planetary ball milling machine (pulversite-6), where the ball-to-material mass ratio was maintained at 6:1 inside a tungsten carbide (45 ml) bowl and milling were continued at 300 rpm for 80 hrs. in air. The milling was intermediately stopped in every 2-hour interval for mixing. The alloyed powders were made into pellets (ϕ ~ 10 mm, *t* ~ 0.3 mm), and heat treated separately at 800 °C and at 1100 °C in air for 24 hrs. to investigate the thermal induced changes in structural phase stabilization and properties. The used samples were denoted as Fe20-XScX_*AY*, where X = 2, 5, 10 for x = 0.2, 0.5 and 1.0, respectively, and *Y* represents (8 for 800 °C and 11 for 1100 °C) the heat treatment temperature.

*2.2. Characterization and measurements*

The X-ray diffraction (XRD) patterns of the samples were recorded by using a RIGAKU X-Ray diffractometer (Cu-Kα, λ =1.5406Å) in the 2θ range 20-80 ° at step size 0.01 °. The neutron diffraction (ND) patterns in the temperature range of 10-300 K were recorded at Dhruva reactor, Bhabha Atomic Research Centre (BARC), Mumbai by using the Powder diffractometer- II (λ = 1.2443 Å). The powdered material (nearly 4-5 g) was loaded in a Vanadium-cylindrical can. The dc magnetization (*M*) of the samples was recorded using Physical Properties Measurement System (PPMS, EC2-Quantum Design, USA). The zero field cooled (ZFC) and field cooled (FC) modes were used to measure the temperature (5-300 K) dependence of magnetization (M(T)) curves. In ZFC mode, the sample was cooled from 300 K to low temperature (5 K) in the absence of external magnetic field (*H*). Then, dc magnetization was measured in the presence of set magnetic field while the sample was warming up to 300 K. In FC mode, the sample was cooled from 300 K under a constant magnetic field down to 5 K and magnetization was recorded while the sample was warming up to 300 K without removing the cooling field. The magnetic field (±70 kOe) dependent magnetization [M(*H*)] curves was recorded at 5 K and 300 K by using ZFC mode and at 5 K by using FC mode where the samples was cooled from 300 K to 5 K in the presence of external field 70 kOe before recording the M(H) loop. The temperature (20-300 K) dependence of electronic properties (conductivity (σ) and current-voltage (I-V) curve) were studied by placing the pellet-shaped sample between the top and bottom electrodes and silver paint was used for good electrical contact. The dc current (I) was measured at constant bias voltage 20 V by using the Keithley 6517B high resistance meter. The electric polarization of the pellets was measured by using a Precision Premier II ferroelectric loop tracer (Radiant tech., USA) and the driving voltage amplitude and frequency were varied depending on measurement temperatures. The low temperature range (20–300 K) was maintained using close cycled cryostat (CCR:CH-204-N 6.5K, Cryo Industries, USA). The magneto-electric (ME) response (the induced voltage across the samples) was measured using a Lock-in-Amplifier (model SR830) in differential voltage mode. The sample was excited under a coaxially ac magnetic field by using a Helmholtz coil pair, which was sourced by an internal ac generator of the Lock-in-Amplifier, and a superimposed dc magnetic field using an electromagnet (MicroSense, USA). The magneto-conductance (MC) was studied by measuring the current at 20 V and varying the magnetic field up to 8 kOe.

## 3. Result and discussion

### *3.1. Low temperature lattice structure and magnetic structure using neutron diffraction pattern*

The neutron diffraction (ND) patterns at different temperatures (≤ 300 K) were refined to get the information of lattice structure (nuclear structure) and magnetic structure. The nuclear structure for ND pattern at 300 K and derived parameters (atomic positions, lattice parameters, cell volume, phase fraction, chemical composition) are consistent to the values obtained from Rietveld refinement of the room temperature XRD pattern [16]. The samples of composition $Fe_{1.8}Sc_{0.2}O_3$ formed single-phased rhombohedral structure with space group $R\overline{3}c$ ($\alpha$-$Fe_2O_3$ phase). The composition $Fe_{1.5}Sc_{0.5}O_3$ (x = 0.5) and $FeScO_3$ (x = 1.0) showed mixed-pattern of $\alpha$-$Fe_2O_3$ -type phase and $Sc_2O_3$ -type cubic phase (*Ia-3* space group). The $\alpha$-$Fe_2O_3$: $Sc_2O_3$-type phase ratio was found ~ 73:27, 94:06, 43:57 and 15:85 for Fe15Sc5_A8, Fe15Sc5_A11, Fe10Sc10_A8, and Fe10Sc10_A11 samples, respectively. The $\alpha$-$Fe_2O_3$-type phase was increased by increasing the heat treatment to higher temperature (1100 $^0$C) for x = 0.5, unlike the increase of $Sc_2O_3$-type phase for x = 1.0 composition.

The same space groups of $\alpha$-$Fe_2O_3$ - and $Sc_2O_3$ -type phases were used for the refinement of magnetic structure. The spin structure between adjacent planes was modelled with *A*-type AFM order, where the Fe spin moments followed ↑↓↓↑ (AFM) sequence along the (001) direction (c-axis) and parallel (either ↑↑ or ↓↓) sequence in the *a-b* planes of α-$Fe_2O_3$ –type phase [20]. Orava et al. [21] modelled the transitions of magnetic space group in α-$Fe_2O_3$ from R $\bar{3}$c1$'$→ C2/c (C2$'$/c$'$) at the Neel temperature $T_N$ ~ 950 K (where spins oriented in the (*a-b*) planes with finite canting among the AFM coupled inter-planar spins, leading to a weak FM/canted AFM (CAFM) structure and ferroelastic properties), and C2/c (C2$'$/c$'$) → R$\bar{3}$c at the Morin transition ($T_M$ ~ 260 K), below which $Fe^{3+}$ spins form out of plane (along *c* axis) collinear AFM structure. The magnetic structure in Sc doped α-$Fe_2O_3$ –type phase is fitted with magnetic space group C2'/c' at all measurement temperatures, confirming its CAFM spin order down to low temperatures [20, 22]. The refinement of propagation vector (*k*) was suitably done from the "k-search" or "WinPlotr-2006" option in the FULLPROF software after providing details of nuclear structure. The ND pattern was best fitted with Pseudo-Voigt peak shape and optimized different fit parameters (scale factor, back ground, atomic positions, lattice parameters ($a$ = $b$ and $c$), FWHM parameters ($u$, $v$ and $w$), Lorentzian isotropic strain parameter (X), profile shape parameter, isotropic thermal parameters ($B_{iso}$), preferred orientation parameters, net moment *M* and magnetic moment components ($M_x$, $M_z$). In polycrystalline sample, the y- components ($M_y$) are averaged out to zero and x-component ($M_x$) defines the *in-plane* moment. The refined ND patterns at 300 K and 10 K/30 K are shown in Fig. 1 (a-j). The intensity variation of the first two peaks (003) and (101), in the insets of Fig. 1, provides the temperature effect on the variation of in-plane magnetic spin structure.

The supplementary Table S1 summarizes the final refinement results of nuclear structure at 300 K and 10 K/30 K. The chemical composition of the samples matched to expected value and remains nearly invariant at different temperatures. The lattice parameter ($c$) in the α-$Fe_2O_3$ phase is found to be 2.71-2.73 times in comparison to in-plane lattice parameter ($a$). The increase of unit cell parameter "*a*" with temperature in Fig. 2 (a-e) and in the insets of Fig. 2 (a-e) showed an overall thermal activated lattice expansion in α-$Fe_2O_3$ – and $Sc_2O_3$ –type phases. Table 1 summarizes the values of Fe-O bond length and <Fe-O-Fe and <O-Fe-O bond angles for the α-$Fe_2O_3$ phase. The Fe-O bonds at the octahedral coordination showed two groups of lengths (e.g. ~ 2.1180 Å and 1.9490 Å, respectively for Fe18Sc2_A8 sample). The increase of average <Fe-O> bond length with temperature is consistent to thermal induced lattice expansion. The average (<Fe-O>) bond length of Sc doped samples with heat treatment at 800 $^{0}$C has decreased with the increase of Sc content. On the other hand, the average bond length has noticeably increased for heat treatment at 1100 $^{0}$C. A proper stabilization of $Sc^{3+}$ ions (radius 0.0745 nm) at the sites of $Fe^{3+}$ ions (radius 0.0645 nm) increased the lattice parameter, which is largely affected by the chemical hybridization of Fe(3d) and O(2p) orbitals. The lattice parameter/Fe-O bond length expansion along the *a-b* planes reduces Fe-O-Fe (AFM) interactions, whereas lattice expansion along the *c-axis* increases the AFM superexchange interactions [22]. Most of the samples showed <Fe-O-Fe bond angles around 94 $^{0}$ and 131.5$^{0}$, and the <O-Fe-O bond angles around 90.5 $^{0}$, 102.8 $^{0}$ and 162 $^{0}$. In Fe10Sc10_A11 sample, the <Fe-O-Fe bond angles have decreased to 91.4 $^{0}$ and 126.8$^{0}$, respectively. In this sample, volume fraction of the non-magnetic $Sc_2O_3$ –type phase has strongly affected the chemical bonding in α-$Fe_2O_3$ system, showing the decrease of <O-Fe-O bond angles to 88.5 $^{0}$ and 101 $^{0}$, and increased the other <O-Fe-O bond angle to 167.5 $^{0}$. According to Goodenough-Kanamori rules [23], the bond angles (<Fe-O-Fe) between 120-180 $^{0}$ prefer a strong AFM superexchange interactions, whereas the bond angles smaller than 120 $^{0}$ and around 90 $^{0}$ support for weak FM superexchange interactions. The results suggest a mixture of distorted FM (*in-plane*) and AFM (*out of plane*) interactions. The dominance of *in-plane* spin order over the *out of plane* spin order results in overall enhancement of FM interactions in Sc doped α-$Fe_2O_3$ system.

Fig. 3 (top panel) demonstrates the modified spin order in Sc doped α-$Fe_2O_3$ structure at low temperatures. Main panels of Fig. 3 show the temperature variation of the spin components. The zero z-component ($\mu_z$) of the spin moment at 300 K suggested a nearly collinear in-plane AFM spin order. A rapid increase of the $\mu_z$ value at temperatures ≤ 250 K indicates canting of the spin structure away from the *a-b* plane, as expected below the Morin transition around 260 K. The non-zero *in-plane* component ($\mu_x$) in the entire temperature range suggests that the spins do not fully reorient along *c*-axis even at 5 K. A similar temperature variation of the integral area of the (003) ND peak to $\mu_x$ components confirms its origin from the *in-plane* magnetic spin order. In case of single-phased Fe18Sc2_A8 sample (Fig. 3(a)), the $\mu_x$ at 300 K is ~ 3.96 μB (≈ μ) and starts to decrease rapidly below 250 K (close to the Morin temperature) with a levelling off at the lower temperatures (~ 2.75 μB). The $\mu_z$ showed a rapid increment below 250 K and saturated at low temperatures. This confirms the breaking of collinear AFM spin order (along *c*-axis) in the Sc doped α-$Fe_2O_3$ system below its Morin transition [1, 24], in contrast to a perfect AFM spin order along the c-axis for un-doped α-$Fe_2O_3$ bulk system [22]. The net moment per $Fe^{3+}$ ion in single phased Fe18Sc2_A8 sample is slightly reduced due to doping of non-magnetic $Sc^{3+}$ ions. Interestingly, neither $\mu_x$ approaches to zero nor $\mu_z$ approaches to net moment (μ ~ 3.90 μB) at low temperatures. The temperature variation of magnetic components in Fe15Sc5_A8 (Fig. 3(b)) and Fe10Sc10_A8 (Fig. 3(c)) samples are identical to that in Fe18Sc2_A8, except the net moment per $Fe^{3+}$ ion is slightly higher for higher Sc content (μ ~ 4.38 μB at 300 K and 4.30 μB at 10 K for Fe15Sc5_A8 sample, and μ ~ 4.68 μB at 300 K and 4.49 μB at 10 K for Fe10Sc10_A8 sample). The $\mu_z$ in these two samples saturates close to the net moment at low temperatures and the $\mu_x$ component settles at lower value (~ 1.50 μB). This shows that Fe moments in Fe15Sc5_A8 and Fe10Sc10_A8 samples are more tilted towards the *c*-axis. In Fe15Sc5_A11 (Fig. 3(d)) and Fe10Sc10_A11 (Fig. 3(e)) samples, the $\mu_z$ slightly increased at low temperatures (0 to ~ 0.50 μB at 10 K) and the $\mu_x$ nearly matched to the net moment M throughout the temperature range. The in-plane moment ($\mu_x$) largely controls the net magnetic moment (μ) of the Sc doped samples with heat treatment at 1100 °C. A local maximum at about 150 K (~ 3.47 μB) followed by a dip at 100 K are marked before making a low temperature magnetic moment upturn in Fe15Sc5_A11 sample, whereas the moment in Fe10Sc10_A11 sample monotonically increased at lower temperatures. The moment/Fe ion values in Sc doped samples are found to be less than the typical paramagnetic spin moment (5 μB) of the $Fe^{3+}$ ions. The decrease of magnetic moment (μ) in the samples with higher heat treatment can be affected by the modified chemical bonding.

*3.2. Low temperature dc magnetic properties*

Bulk magnetic properties of the polycrystalline samples were studied from magnetization curves. The temperature dependence of magnetization curves (Fig. 4 (a-c, e-f)) at 200 Oe showed typical features of non-magnetic Ga doped α-$Fe_2O_3$ system, where breaking of the AFM Fe-O-Fe superexchange bonds has enhanced FM spin order [2]. The peak of dMZFC(T)/dT curves (Fig. 4 (right-Y axis)), irrespective of the ZFC and FC curves, showed Morin transition ($T_M$) at ~ 258 K. The irreversibility between zero-field cooled (MZFC(T)) and field-cooled (MFC(T)) curves in the temperature range 5-350 K and widening of gap at temperatures below 350 K suggest perturbed AFM state. In pure hematite structure, the magnetic spins of $Fe^{3+}$ ions flips from *in-plane* to *out of plane* direction below its $T_M$~260 K. There is practically no magnetic irreversibility below $T_M$ in pure hematite system, where single ion anisotropy ($K_{SI} > 0$) of $Fe^{3+}$ ions dominates at temperatures $< T_M$ and magnetic dipolar anisotropy ($K_D < 0$) between $Fe^{3+}$ moments dominates at temperatures

> $T_M$ [25]. A systematic decrease in the magnetic irreversibility at higher Sc content suggests a decrease of $K_{SI}$ in $Fe_{2-x}Sc_xO3$ system. The modified magnetic properties in Sc doped samples (heat treatment at 800 °C) can be understood from a sharp decrease of the MZFC(T) and MFC(T) curves below $T_M$ and a competitive regime before a making a magnetic upturn below 50 K. The Sc doped $BiFeO_3$ (in rhombohedral structure) [26] and Sc doped $YbFeO_3$ (hexagonal structure) [18] showed similar magnetic upturns below 100 K. However, frustrated AFM state was observed down to low temperature for Sc-doped $LuFeO_3$ (hexagonal structure) system [27]. Such low magnetic upturn in AFM nanoparticles can be explained by formation of core-shell model [28], where AFM spin order is retained inside the core and ground state spin order is perturbed in the shell. The frustrated and uncompensated surface spins contribute to paramagnetic like features at low temperatures. The surface spin frustration is small in single-phased Fe18Sc2_A8 sample. The increasing fraction of $Sc_2O_3$ –type phase in bi-phased samples enhanced the overall low temperature paramagnetic up turn. The FM spin component in the perturbed AFM state has been further improved by applying field-cooled mode (MFC(T) > MZFC(T)) and higher magnetic fields (2-50 kOe). The application of high magnetic field has shifted the Morin transition to lower temperatures (~ 218 K at 40 kOe for Fe15Sc5_A8 sample, 207 K at 50 kOe for Fe10Sc10_A8 sample). The MZFC(T) curves at different fields for Fe18Sc2_A8 sample (Fig. 4(d)) clarified the magnetic field-induced shift of $T_M$ (~ 256 K at 500 Oe to 246 K at 9 kOe) in Sc doped $Fe_2O_3$ system. The presence of Morin transition at higher fields confirmed retaining of a sufficiently strong single ion anisotropy of $Fe^{3+}$ ions, a characteristic feature of the AFM hematite structure, in the Sc doped samples with heat treatment at 800 $^0$C.  The heat treatment of the samples at 1100 °C showed a remarkably modified M(T) curves. The Fe15Sc5_A11 sample (6 % $Sc_2O_3$ phase) showed a signature of Morin transition below 260 K, an overall enhancement of the magnetization and large gap between MFC(T) and MZFC(T) curves at 200 Oe field. A weak magnetic irreversibility between FC and ZFC curves below 260 K shows Morin transition at ~ 256 K and heavy dilution of the AFM spin order in Fe10Sc10_A11 sample, having 85 % of $Sc_2O_3$ phase. The 15 % of $Fe_2O_3$–type FM clusters (ferromagnetism arises from in-plane spin order) in the non-magnetic $Sc_2O_3$ –type matrix showed a strong paramagnetic-type upturn at lower temperatures. A similar feature was seen in Sc doped $YbFeO_3$ system below its AFM ordering temperature at 120 K [27].

The field dependence of magnetization (M(H)) curves at 5 K/10 K and 300 K (Fig. 5) were used to determine bulk magnetic moment of the samples. A clear loop in the M(H) curves and lack of magnetic saturation for fields up to ± 70 kOe confirmed a typical canted AFM/FM structure at all measurement temperatures. The canted AFM features dominated at the low temperature M(H) loop of Fe18Sc02_A8 sample, which showed rod-shaped (elongated) loop with non-linear up curvature in the M(H) curves and high coercivity, whereas the low temperature loop area narrowed down to low coercivity for the samples with higher Sc content (x = 0.5, 1.0) and heat treated at 800 $^0$C. M(H) curves of these samples showed a nearly parallelogram-shaped wide loop and high coercivity at 300 K, which confirmed dominated canted FM features above the Morin transition. M(H) curves of the Fe10Sc10_A11 showed features, which are similar to the samples with heat treatment at 800 $^0$C. The Fe15Sc5_A11 sample showed nearly similar character (wide loop and high coercivity) in the M(H) curves both at 5 K and 300 K. In order check the exchange bias effect (magnetic shift due to exchange coupling between FM and AFM layers), the M(H) loop at low temperatures were recorded for selected samples after field-cooling (in the presence of 70 kOe) from 300 K to the low measurement temperature. A comparison between FC- and ZFC-M(H) plots (insets of Fig. 5) shows a clear shift of the FC loop with reference to the ZFC loop in the samples (Fe10Sc10_A8, Fe10Sc10_A11) with high Sc content and larger fraction of $Sc_2O_3$ –type phase.

On the other hand, the Fe15Sc5_A11 (x = 0.5) sample with heat treatment at 1100 $^{0}$C and small fraction of $Sc_2O_3$ –type phase practically does not exhibit any shift of the FC-loop.

The average of magnetic parameters (coercivity ($H_C$), remanent magnetization ($M_R$)) were calculated from positive and negative axes of the loops. The center ($H_0$ and $M_0$) of the loops was determined from $H_0 = (H_{C1} + H_{C2})/2$ and $M_0 = (M_{R1} + M_{R2})/2$, respectively. The exchange bias field ($H_{exb}$) and magnetization ($M_{exb}$) were determined from shift of the center of FC-loop with reference to the center of ZFC-loop ($H_{exb} = (H_0^{FC} – H_0^{ZFC})$ and $M_{exb} = (M_0^{FC} - M_0^{ZFC})$). The field-cooling induced shift in the coercivity ($\Delta H_c$) and remanent magnetization ($\Delta M_R$) of the FC-loop with reference to the ZFC-loop were calculated from $\Delta H_c = H_C^{FC} – H_C^{ZFC}$ and $\Delta M_R = M_R^{FC} – M_C^{ZFC}$, respectively. The calculated values of the magnetic parameters are summarized in Table 2. The Fe10Sc10_A11 sample showed high amount of exchange bias shift and other magnetic parameters. The presence of $Sc_2O_3$ –type phase has catalysed for a significant reduction of magnetic coercivity in Fe2-xScxO3 system at higher values of Sc (x), especially at 5 K. In consistent to the results of magnetic moment per Fe atom from ND patterns, Table 2 confirms that the secondary $Sc_2O_3$ –type phase helps for a substantial enhancement of bulk magnetization in $Fe_{2-x}ScxO_3$ system and plays significant role in modifying the *in-plane* $Fe^{3+}$-O-$Fe^{3+}$ superexchange coupling in Rhombohedral structure of $\alpha$-$Fe_2O_3$ –type phase. The off-plane spin canting angle ($\theta$) values of the $Fe^{3+}$ spins at 5 K and 300 K were calculated by using the formula $\theta = sin^{-1}(M/\mu_x)$, where M is the magnetic moment from M(H) curves at 50 kOe and $\mu_x$ is the in-plane moment from the refinement of ND patterns. The results in Table 3 suggest that off-plane canting angle ($\theta$) slightly increased at low temperatures, but it never reached to 90 $^{\circ}$; a case for perfect collinear AFM spin order in hematite structure along the *c*- axis. The higher $M_R$ at 300 K than the values at 5 K confirmed that *in-plane* FM spin order still dominates at temperatures above the Morin transition and AFM back ground is sufficiently strong at low temperatures, except the paramagnetic contribution dominates at low temperatures for the Fe10Sc10_A11 sample having 85 % of $Sc_2O_3$-type phase.

3.3. Low temperature electrical conductivity and polarization

The temperature dependence of dc current $I_{dc}(T)$ at dc bias voltage ($V_{dc}$) = 20 V was measured during warming of the samples from 25 K to 300 K (@ 0.5-1.0 K/ minute and 1.5 K interval). The electrical dc conductivity was calculated by using the formula $\sigma_{dc}(T) = I_{dc}(T) d/A V_{dc}$. The sample was waited in the presence of $V_{dc}$ for 10 s before measuring $I_{dc}$ at specific temperature. Then, $V_{dc}$ was made OFF during sweeping the temperature to next set point. The $\sigma_{dc}(T)$ curves in Fig. 6(a) confirmed a transition from an insulating (conductivity ~ $10^{-14}$ -$10^{-10}$ S/cm) state to a high conductive (conductivity ~ $10^{-10}$ -$10^{-7}$ S/cm) state as the measurement temperature reaches above Morin transition (260 K). The temperature dependence of electrical charge density (Q/area) curves measured at $V_{dc}$ = 20 V (Fig. 6 (b)) also confirmed a transition from insulating state to high conductivity state above the Morin transition. The charge measurement stopped at higher temperatures when crossed the measurement limit of 2 μC for Keithley 6517B meter. The electrical charge density was found typically 88-475 nC/cm$^3$ at 200 K and then started to increase sharply as the temperature increases above 240 K (depending on samples). The samples with higher Sc content and higher heat treatment temperature showed substantially high electrical charge density. The Fe15Sc5_A11 sample showed the highest charge density and Fe18Sc2_A8 sample showed the least charge density. A strong correlation between the electrical conductivity and change of magnetic spin order (canted AFM state to canted FM state above 260 K) was theoretically [29, 30] predicted in hematite structure, where enhancement of the conductivity was explained in terms of small polaron (bound state of electrons) formation at the defect sites and anisotropic nature of the charge hopping via the $Fe^{3+}$-O-$Fe^{2+}$ superexchange paths in hematite structure. The electronic charge hopping between Fe-O-Fe sites along *c*-axis and AFM spin order prefer an insulating state at temperatures below Morin transition. The canted-FM spin order and charge hopping via Fe-O-Fe superexchange paths along (*a-b*) *in-plane* directions can enhance electrical conductivity at temperatures above Morin transition nearly four times higher than the values along *out of plane* direction at low temperatures. The conductivity data in the low temperature and high temperature were fitted with Arrhenius law $\sigma(T) = A\, e^{-\frac{E_a}{k_B T}}$ [Fig. 6 (c-g)] with activation energy ($E_a$) in the range of 0.04-0.14 eV and 0.52 – 2.21 eV, respectively. The activation energy in the low temperature suggest polaron hopping within a distance of 40 nm and higher activation energy at the higher temperature is most probably affected by the Fe-site vacancy [29] due to segregation of a fraction of lattice structure into $Sc_2O_3$-type phase. The Fe10Sc10_A11 sample (contained 85 % of $Sc_2O_3$-type phase) showed the highest activation energy (2.21 eV) and the Fe15Sc5_A11 sample (contained 6 % of $Sc_2O_3$-type phase) showed the lowest activation energy (0.52 eV). The interaction potential between the polarons and lattice defects has been modified by substituting $Sc^{3+}$ ions at the $Fe^{3+}$ sites, and resulted in considerable increase of the electrical conductivity down to low temperature. The extension of the in-plane (canted) FM spin order and increase of conductivity down to low temperature in Sc doped hematite structure opens a wide scope for tuning magnetic and electronic states over a wide temperature range [31, 32].

The voltage dependence of electric polarization (P(V)) curves were measured at selected temperatures (50-300 K). The amplitude of a triangular shaped driving voltage was suitably varied from 100 V to 1 kV and frequency varied in the range 16.67 Hz to 1 kHz to avoid polarization leakage. We observed that conductivity effect dominates at temperatures above 280 K. The P(V) curves in Fig. 7 (a-d) within voltage range 100 V/500 V showed a linear capacitance behaviour with minor loop at temperatures below 250 K and ferroelectric polarization-like loops above 250 K. The maximum polarization ($P_{max}$) in the samples was found in the range of 0.5-8 μC/cm$^2$ with remanent polarization ($P_r$) in the range of 0-2 μC/cm$^2$ and coercive voltage ($V_C$) in the range of 0-300 V. The maximum polarization ($P_{max}$) was normalized by maximum driving voltage ($V_{max}$) for uniformity in presentation. Fig. 7(e-g) show temperature variation of $P_{max}/V_{max}$ (capacitance per unit area), $P_r/P_{max}$ (squaresness ratio) and coercive voltage ($V_C$). The $P_{max}/V_{max}$ and $V_C$ showed either a peak or a rapid increase in the temperature range of 250-270 K. The results suggest that electric polarization is sensitive to the process of spin flipping from canted AFM state (out of plane) at lower temperatures to canted in-plane FM state above the Morin transition. In the spin flipping process the anisotropic nature of spin exchange interactions can break inversion symmetry due to Dzyaloshinskii-Moriya interaction DMI = $(\overrightarrow{D_{ij}}.\left(\overrightarrow{S_i} \times \overrightarrow{S_j}\right)$, which is favourable to produce ferroelectric-like polarization in α-$Fe_2O_3$-type systems [33, 34]. The Ga doped $Fe_2O_3$ system also showed a peak in relative dielectric permittivity and ac conductivity near to spin flipping zone [2, 3, 35]. The P(V) loops at room temperature are found to be influenced by the conductivity effect. The intrinsic ferroelectric-like polarization in the samples at room temperature was cross-checked from measurements of PUND and P(V) loops at different frequencies. The PUND response was measured (at pulse amplitudes 10 V and 50 V) for single-phased Fe18Sc02_A8, and bi-phased Fe15Sc05_A8 and Fe10Sc10_A8 samples. A difference between the polarization during switching pulse (polarization + contribution from conductivity and dielectric loss) and non-switching pulse (contribution from conductivity and dielectric loss)) in both positive and negative pulse directions (see in supplementary Fig. S1(a-c)) confirmed non-zero value of the switchable polarization. The PUND results for Fe15Sc05_A8 sample at 10 V can be summarized as dP(+) = 0.00378, dPr (+) = 0.00548, dP (-) = - 0.00368 and dPr (-) = -0.00548 in μC/cm2 unit. On the other hand, good quality P(V) loops without leakage (in Fig. 7 (h)) were measured at pulse amplitude of ± 1 kV and frequencies (8-33.3 Hz) for Fe18Sc02_A8 sample after cooling the sample from 425 K to 300 K under electric poling of voltage 2.5 kV. The electric coercivity ($V_C$) increased with frequency ($f$) and its scaling according to $V_C \sim f^{\beta}$ with $\beta \sim 0.45(1)$ (Fig. 7 (i)) support the existence of non-zero intrinsic polarization in materials [36]. A minor decrease of the $P_{max}$ with the increase of frequency and later on tending to level off at higher frequencies (Fig. 7(j)) suggest small contributions from Maxwell-Wagner polarization and electrode/interface at the lower frequencies.

The I-V curves within ± 20 V at 250 K, 280 K and 300 K (Fig. 8 (a-e) guide their temperature evolution. The bias voltage cycling does not show significant I-V loop features at temperatures ≤ 250 K. The loop features at higher temperatures (e.g., 300 K in Fig. 8 (f-j)) are characterized by a typical transition from high resistance state (HRS) during the bias voltage sweep + 20 V → 0 V (path 1) to low resistance state (LRS) during the bias voltage sweep + 0 V → -20 V (path 2) to high resistance state (HRS) during the bias voltage sweep -20 V → 0 V (path 3) and low resistance state (LRS) during the bias voltage sweep + 0 V → + 20 V (path 4). The HRS/LRS-like states are consistent with resistive switching in the materials and they could be associated with lattice defects and polaron formation at the atomic scale [31]. The I-V curves at 300 K indicated a coexistence of polarization current due to switchable electric polarization and leakage due to conductivity effect. The coexistence of both ferroelectric-like polarization and high conductivity state in the same temperature range above the Morin transition makes the isolation of the electric polarization from the conductivity effect difficult. In such cases, time-integration of the current-voltage (I-V) curves is a useful technique to subtract the conductivity contribution and extract the ferroelectric-like polarization [37]. The electric polarization at 300 K was, first, calculated from the time integration of the recorded I-V curves using the formula P = $\frac{\int_{path\,1}^{path\,4} I(t)dt}{A}$. The calculated P(V) curves (red curves), as shown in Fig. 8 (k, l, n), clearly accompanied with leakage component. Theoretically, a perfect ferroelectric polarization curve (without leakage) should provide current values near to zero during path 1 (saturation polraization at positive side) and path 3 (saturation polraization at negative side). The increasing I(V) curves during path 1 and path 3 were estimated as the amount of leakage currents at the positive and negative sides of bias voltages, respectively. The I(V) curves in paths 1 and 3 were fitted with polynomial function of order 3. The fitted data were subtracted from the experimental I(V) data (red curves in Fig. 8 (f-j)) at the positive and negative sides of the bias voltage, respectively. The resultant I(V) curves (blue curves in Fig. 8 (f-j)) were used to calculate the switchable polarization $P_S = \frac{\Delta t}{\Delta V}\frac{\int_{path\,1}^{path\,4} I(V)dV}{A}$. The $P_S$(V) curves (blue colour), as shown in Fig. 8 (k, l, n), suggest switchable polarization ($P_{max}$) in the range of 4-16 μC/cm$^2$ and electric coercivity ($V_C$) in the range of 8.5-9 V. The estimated switchable polarization values in our samples are within the experimental value ~5 μC/cm$^2$ in $ScFeO_3$ [14], although well below of the theoretically predicted maximum ~ 105 μC/cm$^2$ in polar phase (R3c) of $ScFeO_3$ [10].

3.4. Charge-spin coupling effect

A possible electronic charge-spin coupling effect in Sc doped hematite samples has been verified from the measurement of magneto-electric (ME) voltage ($V_{ME}$) and magneto-conductance (MC) across the samples. The in-phase and out of phase components of the $V_{ME}$ were measured by Lock-in-Amplifier using the methods of frequency sweep of an ac magnetic field at amplitude 2.64 Oe, along with superimposed dc magnetic field ($H_{dc}$) of 200 Oe, and dc magnetic field sweep (0 to 8 kOe) in the presence of an ac magnetic field at amplitude 10 Oe and frequency 1997 Hz. The $V_{ME}$ was measured in transverse geometry, where the sample plane direction (z-axis) was kept perpendicular to in-plane (x-y) magnetic field direction (x-axis) to minimize the electromagnetic induction effect [38, 39]. The characteristics of the total ME voltage curves were largely dominated by in-phase component ($V_{ME}^{/}$) and out of phase-component ($V_{ME}^{//}$) was small at lower frequencies. The $V_{ME}^{//}$ value, which includes contributions from inductive pickup (electromagnetic induction) and magneto-impedance effects, starts to dominate at frequencies typically above 6.5-7 kHz (see in supplementary Fig. S1(d)). The ME coupling coefficient ($\alpha_{ME} = \frac{1}{t}\frac{V_{ME}^{/}}{H_{ac}}$) was calculated from the $V_{ME}^{/}$ response, whose origin is largely associated with intrinsic magnetostriction and spin-induced electric polarization [5, 35]. The frequency scan of $V_{ME}^{/}$ showed a non-linear increase, similar to the pattern of $\alpha_{ME}$ variation in Fig. 9 (a). The maximum $\alpha_{ME}$ (peak) values (4.10-32.80 mV/Oe.cm) from the frequency scan were obtained correspond to the maximum of $V_{ME}^{/}$ values ~ 0.37- 2.7 mV at the frequency ~ 6 kHz. The dc magnetic field dependence of MEV curves in Fig. 9(b-c) were normalized with their initial values at 50 Oe for better presentation. The normalization values are shown inside the graphs. The dc field scan of $V_{ME}^{/}$ (Fig. 9(b)) shows a non-linear responses. The dc field range of 3.7-5 kOe, where $V_{ME}^{/}$ exhibited a broad maximum, falls within the magnetic field range where the samples exhibited canted ferromagnetic M(H) loop features (magnetic irreversibility). The $V_{ME}^{/}$ response gradually decreased at higher dc fields, where M(H) curves showed reversibility and linearity due to weakening of domain-wall motion of the canted spin structure. On the other hand, total ME voltage in Fig. 9 (c) showed nearly a linear increase with dc magnetic field, where dc fields act as the bias during ME response under excitation of ac magnetic field at amplitude 10 Oe and frequency 1997 Hz. A few multiferroic materials, e.g., bi-layer composite of single crystal $SrAl_xFe_{12-x}O_{19}$ hexaferrites and PbZrTiO3 ceramics, showed a linear response of MEV with dc magnetic field. Such materials can be useful as magnetic field sensor [40]. Otherwise, multiferroic systems generally showed a non-linear increase in MEV vs $H_{dc}$ curve and a peak at certain dc magnetic field. The $V_{ME}^{/}$ curve showed similar response in our samples. The peak of $V_{ME}^{/}$ curve (3.71 mV) for the single-phased Fe18Sc02_A8 sample matched to its peak in total MEV at dc bias field of 7.4 kOe. Otherwise, total MEV values (0.33-5.05 mV) at 8 kOe of the bi-phased samples were found relatively high in comparison to the peak values (0.19-3.54 mV) in the dc field dependence of $V_{ME}^{/}$ curves at dc field around 5 kOe. The additional contributions in the total MEV at higher dc fields is understood from the out of phase signal. The peak values of $V_{ME}^{/}$ ~ 0.19- 3.71 mV correspond to $\alpha_{ME}$ values ~ 0.62-10.11 mV/Oe.cm in our samples.

The magneto-conductivity (MC) effect was studied by measurement of the current passing through the samples at bias voltage 20 V and dc magnetic field sweep from 0 to 8 kOe and back to 0 Oe. The I(H) curves were measured in longitudinal geometry (normal of sample plane, measured current flow and applied voltage, and dc magnetic field are in same direction (x-axis), for maximum exposure of the sample (surface) magnetization to magnetic field (flux = $\vec{H}.\vec{A}$). The I(H) curves (in Fig. 10) showed a rapid decrement of current on increasing the magnetic field (H) and later on nearly saturated as the field approaches to 8 kOe. During reversing of the magnetic field from 8 kOe to 0 Oe, the current value showed a minor decrement from the value at 8 kOe in most of the samples, except an increase of the current while reversing back from 8 kOe for the Fe10Sc10_A8 sample, thereby exhibiting irreversible current paths in all the samples. The MC behaviour in the canted FM/AFM systems with moderate magnetic coercivity can be understood from a slow domain wall motion under magnetic fields [41]. A remarkably high MC (32-90 %) values, using the formula MC = $\frac{I(8kOe)-I(0)}{I(8kOe)} x100$ (%), support a strong MC effect in the samples.

The nature of lattice and spin structure at the grain-boundaries of single- or multi-phased samples controls the magnetic spin exchange interactions and electron scattering and MC process [42].

3.5. Role of magnetic α-$Fe_2O_3$-type phase and non-magnetic $Sc_2O_3$ –type phase

The roles of α-$Fe_2O_3$-type canted FM/AFM phase and secondary (non-magnetic) $Sc_2O_3$ –type phase have been understood by comparing the ME parameters (MEV at 8kOe and peak value of $V_{ME}^{/}$, and $\alpha_{ME}$ from the $V_{ME}^{/}$ curves in dc magnetic field scan), ferromagnetic parameters ($\mu_{x-y}$, $H_C$), and electrical parameters ($\sigma_{20V}$, $J_{20V}$, $P_{max}$ and $V_C$) of the samples at room temperature with the variation of $Sc_2O_3$ –type phase. The electrical parameters ($\sigma_{20V}$, $J_{20V}$, $P_{max}$, $V_C$) were extracted from I-V curves at room temperature. Fig. 11 (a-f) clarified a reasonably high values of magnetic moment, magnetic coercivity, ME response, MC response, electrical conductivity and electric polarization in single-phased Fe18Sc2_A8 sample. The $Sc_2O_3$ –type phase also played a significant role on tuning the electrical, magnetic and charge-spin coupling properties. However, physical parameters of the system do not vary monotonically with the secondary phase, which coexists with canted FM/AFM α-$Fe_2O_3$-type phase. The sample (Fe10Sc10_A11) with high Sc content and highest secondary phase (85 %) showed ME response, ferromagnetic parameters, electrical polarization and conductivity in the range of least values despite of its highest amount of MC (90 %). On the other hand, Fe15Sc05_A11 sample with the least amount of $Sc_2O_3$ –type phase (6 %) showed the least amount of magnetic and ME parameters in contrary to its highest amount of electrical conductivity and polarization, and high amount of MC (77 %). Our results suggest that secondary phase fraction does not play the primary role. The heat treatment at 800 °C seems to be the optimum condition for achieving the good values of magnetic and magneto-electric parameters, although electrical conductivity moderately decreased, irrespective of the Sc content smaller (Fe15Sc05_A8) or larger (Fe10Sc10_A8). It is understood that coexistence of non-magnetic $Sc_2O_3$ –type phase can act as a bridge to control chemical bonding, magnetic superrexchange interactions and electronic structure by substituting Fe ($3d^5$)-O(2p) covalent bonds with non-magnetic Sc ($3d^0$)-O(2p) bonds. The $Sc_2O_3$ –type phase provides additional electron scattering centers during charge-spin diffusion at the grain boundaries of magnetic α-$Fe_2O_3$ and non-magnetic $Sc_2O_3$ -type lattice structures, which in turn influence the ME and MC properties [11, 12, 43]. The additional contributions in out of phase component due to electro-magnetic induction and magnetoresistance /magnetoimpedance effects were found relatively large in the mixed-phased samples at the higher frequencies (> 6.5 kHz) and higher dc magnetic field bias (>5 kOe). This makes a considerable difference in the values of MEV and coupling constant determined from the in-phase component and total MEV of the samples. Otherwise, a reasonably strong ME response and magnetic field induced polarization are confirmed in single-phased and mixed-phased samples of $Fe_{2-x}Sc_xO_3$ system. The ME coupling constant ($\alpha_{ME}$) values from the in-phase response in our samples fall within the range that reported in bulk multiferroic oxides ($\alpha_{ME}$ ~ 0.5-7 mV/Oe/cm for $BiFeO_3$ and ~ 0.76-1.67 mV/Oe/cm in non-magnetic Mg doped GdFeO3 [44, 45].

As per the traditional concept, ferroelectric polarization is permitted only in polar phase (R3c structure) and no ferroelectric polarization for centrosymmetric R-3c structure of $ABO_3$-type oxides. It may be clarified that XRD or neutron diffraction patterns are identical both in polar R3c phase and non-polar R-3c phase. The difference may be realized from the position of oxygen atoms at z = 0.25 for R-3c phase and its slight off-center for R3c phase [6]. The neutron diffraction patterns of the Sc doped α-$Fe_2O_3$ phase were best fitted with non-polar (R-3c) phase by fixing position of oxygen atoms mostly at z = 0.25. As shown in supplementary table S1 for Fe15Sc5_A8 sample, the position of oxygen atoms may be allowed for a finite amount of off-centering about z

= 0.25. Hence, there may be a possibility of non-zero amount of non-centrosymmetric (R3c) phase at the interfacial or grain boundary regions. At the atomic scale, a fine variation in the atomic position and site occupancy of O and Fe/Sc atoms and lattice disorder, which are not generalized, could be enough for producing ferroelectric polarization in centrosymmetric oxides [8, 9]. Even a minor shift of Li atoms within 0.5 Å along the c-axis was able to produce ferroelectric polar state in electrically metal $LiOsO_3$ [6].

The $Fe_{2-x}Sc_xO_3$ system was prepared by alloying of non-polar $\alpha$-$Fe_2O_3$ and $Sc_2O_3$ oxides. All the samples, irrespective of the single- or bi-phased structure, retained the Morin temperature around 260 K, which is related to the transition of spin order from in-plane to out of plane direction at lower temperatures and controlled by DMI. The refinement of ND patterns confirmed non-collinear spin structure down to 5 K. From magnetic point of view, ferroelectric-like polarization in rhombohedral $ABO_3$ oxide is possible due to non-collinear (canted) spin order where anisotropic nature of DMI break the inversion symmetry [33]. The breaking of low temperature AFM ground state in $\alpha$-$Fe_2O_3$ by Sc doping modifies the out-of-plane ($Dz$) and in-plane ($Dx$, $Dy$) components of the antisymmetric DMI. This can produce meta-stable spin-lattice structure (polar-state) with enhanced *in-plane* FM moment and ferroelectric-like polarization [13 19]. A significant change in structural parameters (lattice expansion, oxygen atom position and its temperature factor (B)) above the Morin transition can take into account of the effects of meta-stable (canted-spin order induced) polar phase. The observation of electric polarization value either maximum near to the spin-flipping zone (260-270 K) or higher value in the temperature range of canted FM state suggest that electrical charge-spin transport properties are sensitive to the process of spin orientation from out of plane to in-plane direction (spin-flipping zone) in non-magnetic Sc ($3d^0$) or Ga ($3d^{10}$) doped $\alpha$-$Fe_2O$ systems [3, 34, 35]. In such material, ME response is driven by canted FM structure induced spin current [46, 47]. The results show that Sc content can be optimized near to 0.5 to achieve the maximum charge-spin coupling, electrical conductivity and ferroelectric polarization in the $Fe_{2-x}Sc_xO_3$ system. Such optimization Sc content for achieving the maximum ME response in the mixed system is consistent to the previous report [48]. The increased FM order, tuning of electrical conductivity, I-V characteristics, electric polarization, and charge-spin coupling in Sc doped hematite system may be useful for making non-volatile electric and magnetic field controlled spintronic and low power energy devices [30, 32, 43, 46-50].

## 4. Conclusions

The rigid magnetic spin structure (collinear AFM spin order below Morin transition ~ 260 K and canted FM spin order above Morin transition) of the $\alpha$-$Fe_2O_3$ structure has been broken by Sc doping. The strategy of breaking AFM ground state of $\alpha$-$Fe_2O_3$ by doping non-magnetic $Sc^{3+}$ ($3d^0$) ions at the $Fe^{3+}$ ($3d^5$) sites played the primary role for a remarkable enhancement of electrical conductivity, ferroelectric-like polarization and charge-spin coupling in Sc doped $\alpha$-$Fe_2O_3$ system. The ND patterns may be well fitted with non-polar phase (Rhombohedral structure with R-3c space group) in macroscopic scale. A possibility of non-zero amount of polar (non-centrosymmetric R3c) phase cannot be ruled out by considering a meta-stable lattice structure, lattice disorder, modified chemical bonds at the interfacial phase or grain boundary regions where the secondary phase may play significant role. The transformation of spin orientation from out of plane to in-plane direction, via Dzyaloshinskii-Moriya interactions, played the crucial role on producing finite ferroelectric-like polarization, strong charge-spin coupling, and a transformation from low magnetic and conductivity state (at low temperatures) to high magnetic and conductivity state at temperatures above Morin transition. The work may open a wide scope for further investigations on tuning the ferromagnetic, ferroelectric, electrical conductivity and magneto-electric properties by controlling the doping content, secondary phase fraction and heat treatment in hematite based doped materials, and brightens their scope towards the multifunctional electronic device applications.

Acknowledgment

The authors acknowledge the research grant (No. 58/14/18/2022-BRNS) from BRNS, Government of India for carrying out this experimental work.

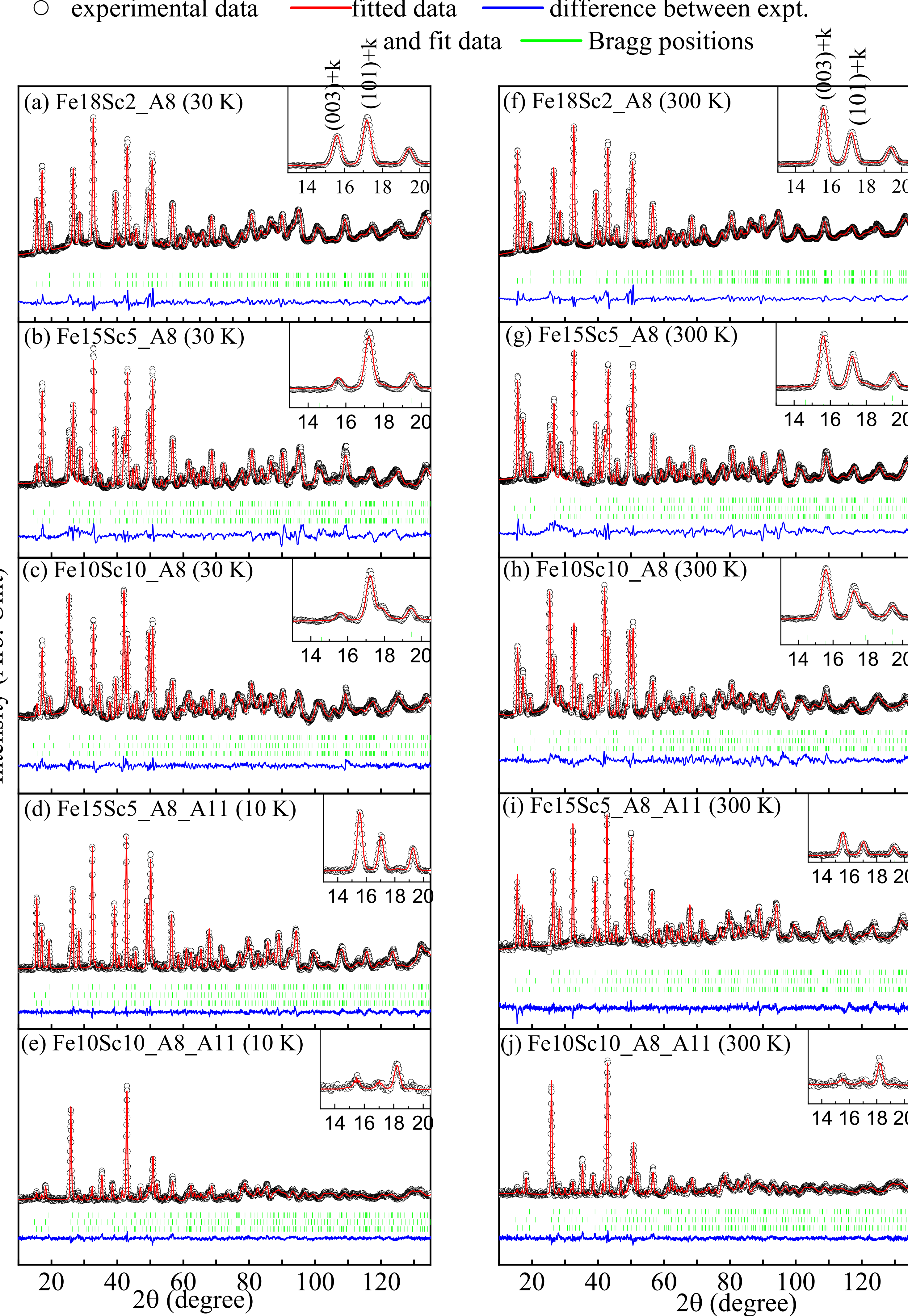

Fig. 1 Rietveld Refined Neutron diffraction patterns of the samples at 10 K/30 K (a-e) and at 300 K (f-j). Insets show the contributions from (003) and (101) planes.

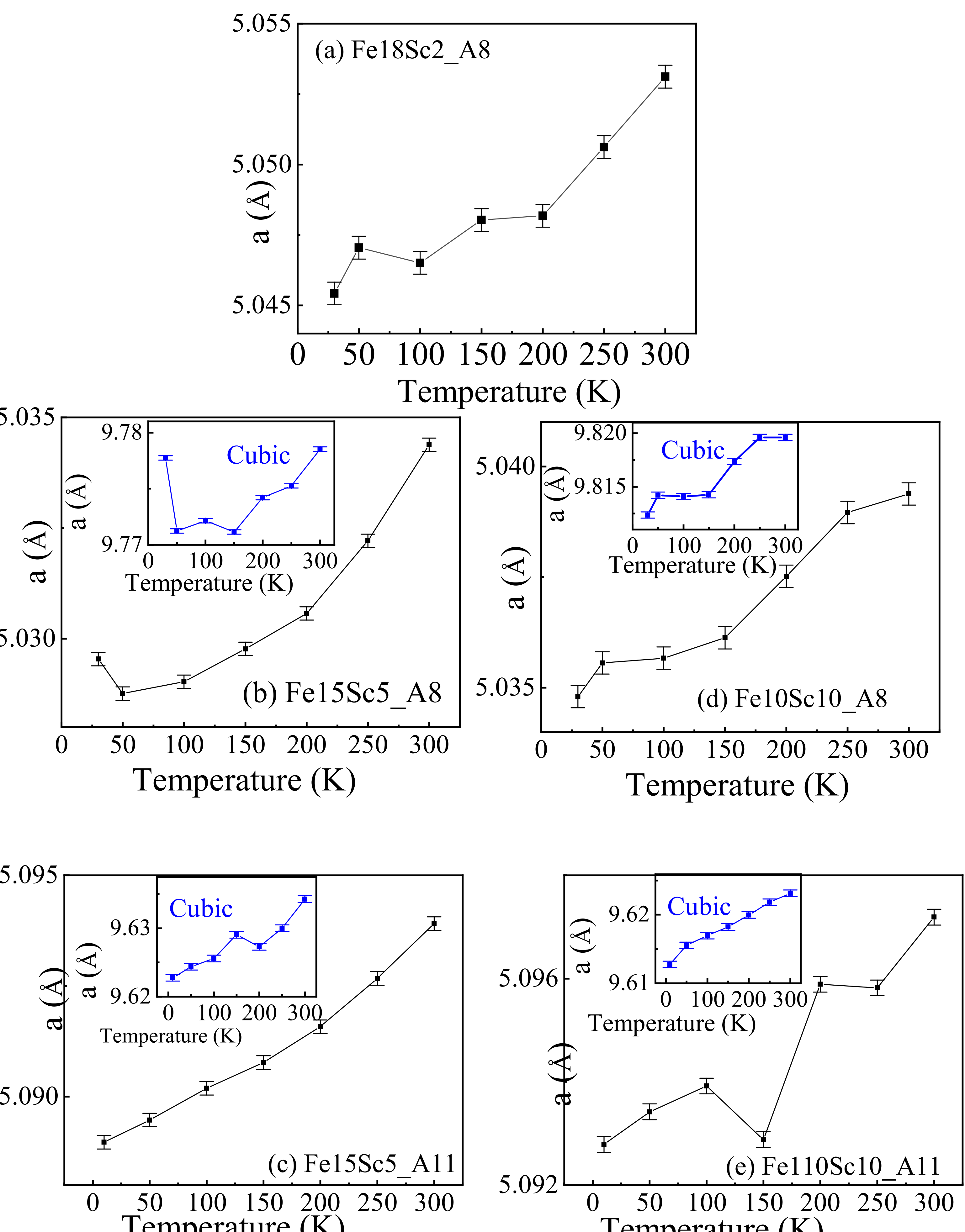

Fig. 2 Temperature variation of the lattice parameters for the Rhombohedral phase (main frame) and the cubic phase (inset) of the samples (a-d).

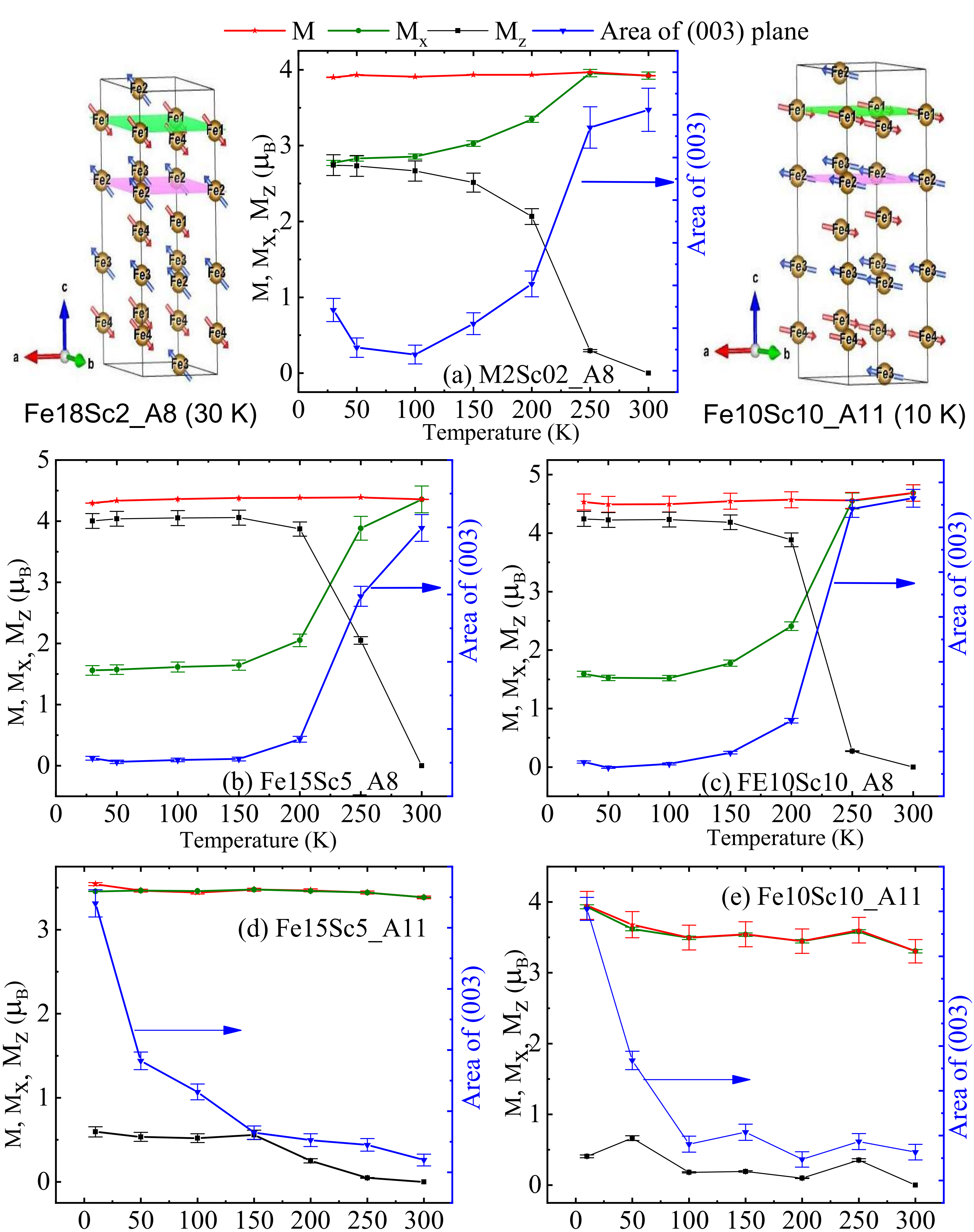

Fig. 3 Temperature dependence of net moment (M), in-plane moment ($M_x$), z-component ($M_z$), and integrated area of the (003) plane for the samples Fe18Sc2_A8 (a), Fe15Sc5_A8 (b), Fe10Sc10_A8 (c), Fe15Sc05_A11 (d), and Fe10Sc10_A11 (e). .

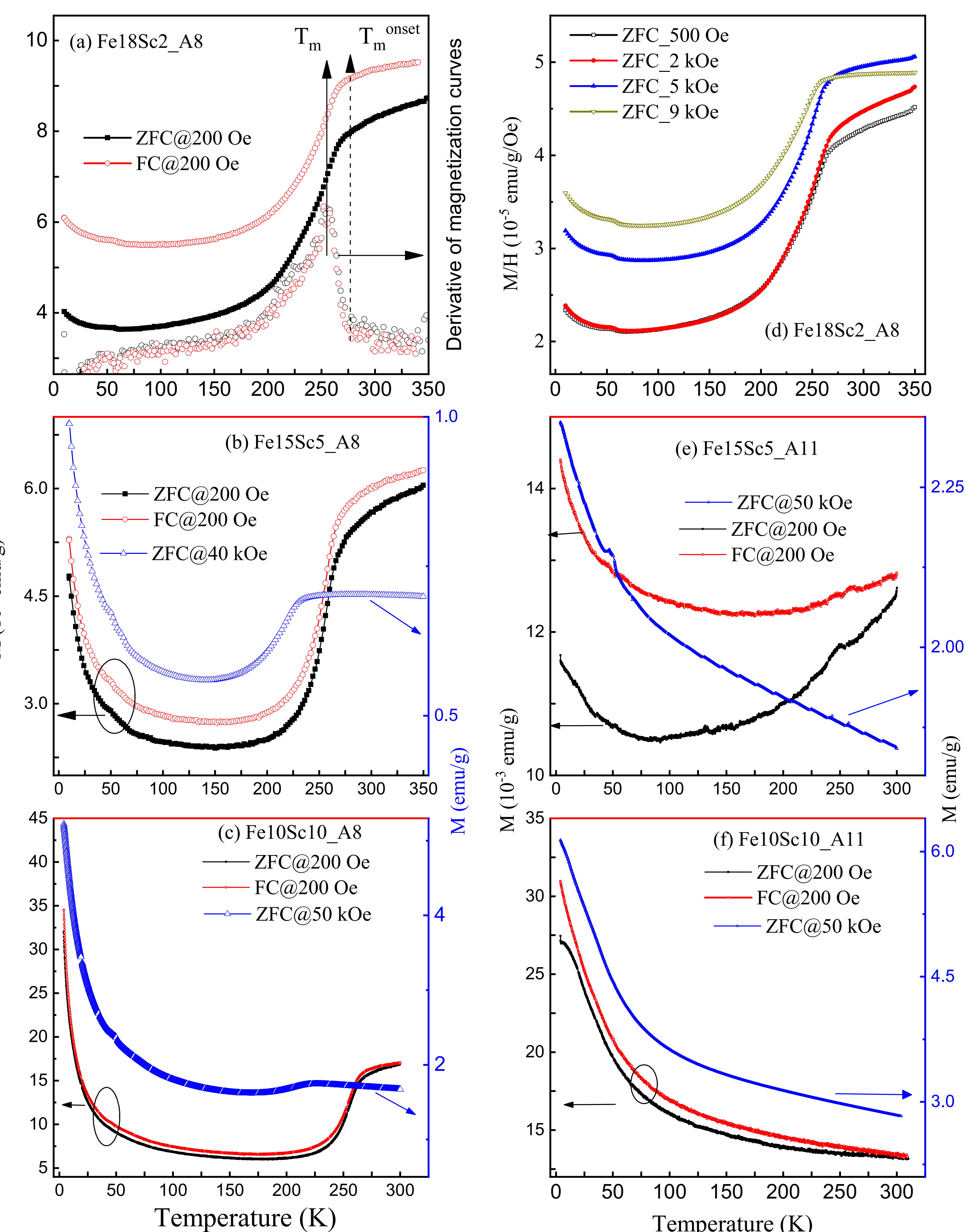

Fig. 4 Temperature dependence of the ZFC and FC magnetization (M) curves at 200 Oe (left-Y axis) and ZFC curves at 40/50 kOe (right-Y axis) of (a, b, c, e, f). The temperature derivative of the magnetization curves shows Morin transition (right Y axis in (a)). (b) The field normalized MFC/H vs Temperature curves at 500 Oe, 2 kOe, 5 kOe and 9 kOe for the Fe18Sc2_A8 sample.

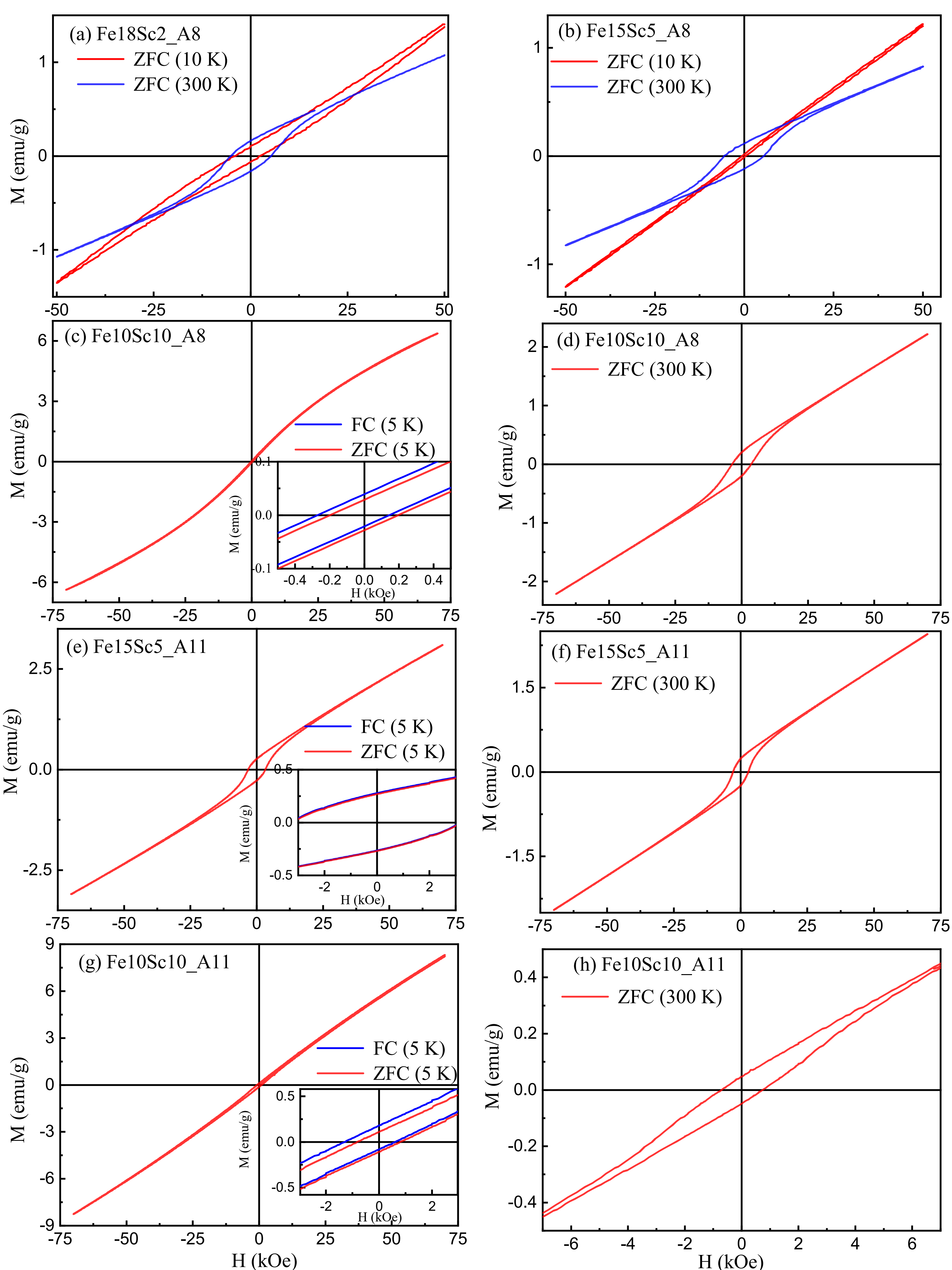

Fig. 5 M-H loops measured at 5 K/10 K and 300 K. The insets clarified the differences between ZFC and FC (@ 70 kOe)- M(H) loops in the small field range.

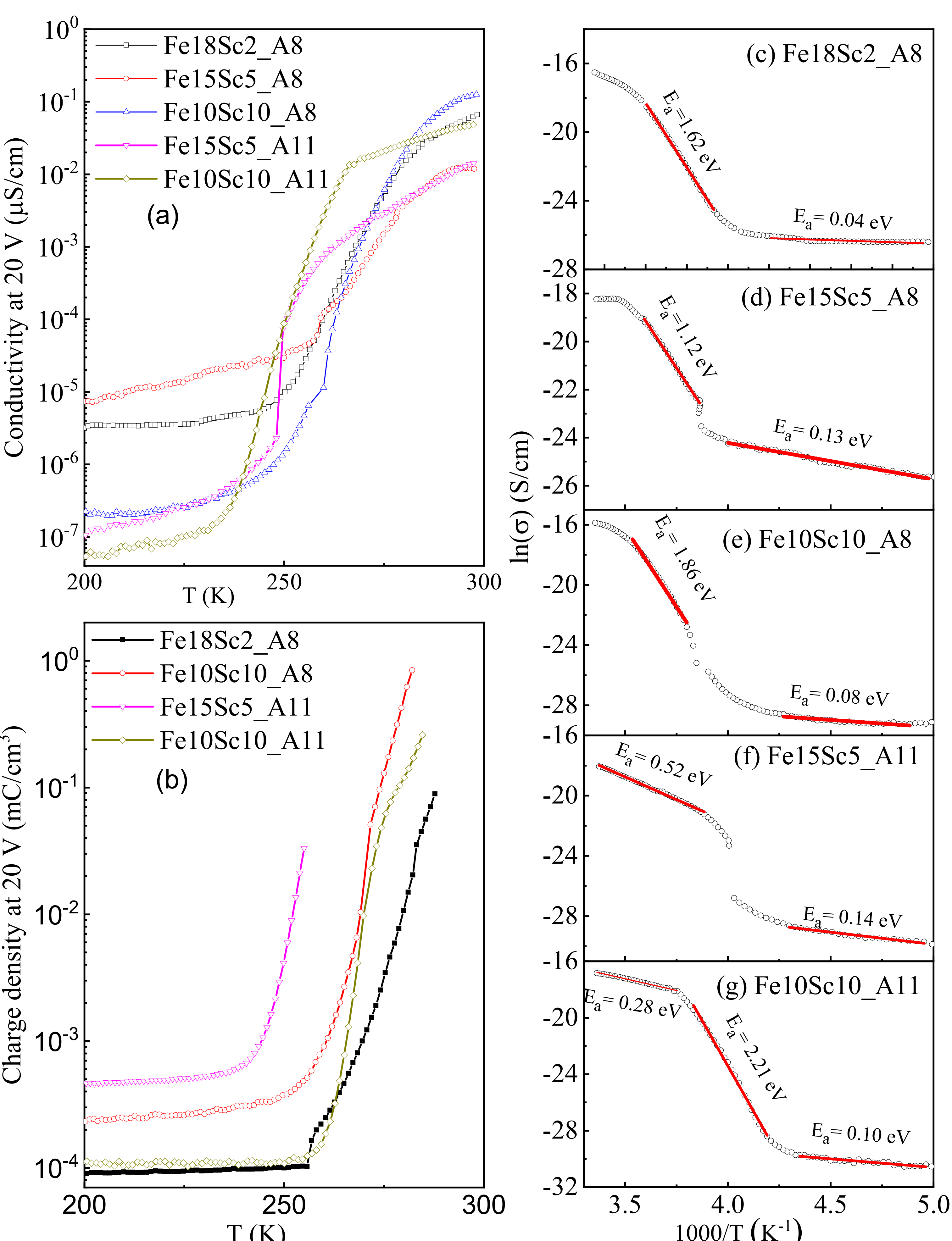

Fig. 6 Temperature dependence of the electrical conductivity (a) and charge density (b) for the samples measured at bias voltage 20 V. (c) The conductivity data fitted using Arhenius law to calculate activation energy.

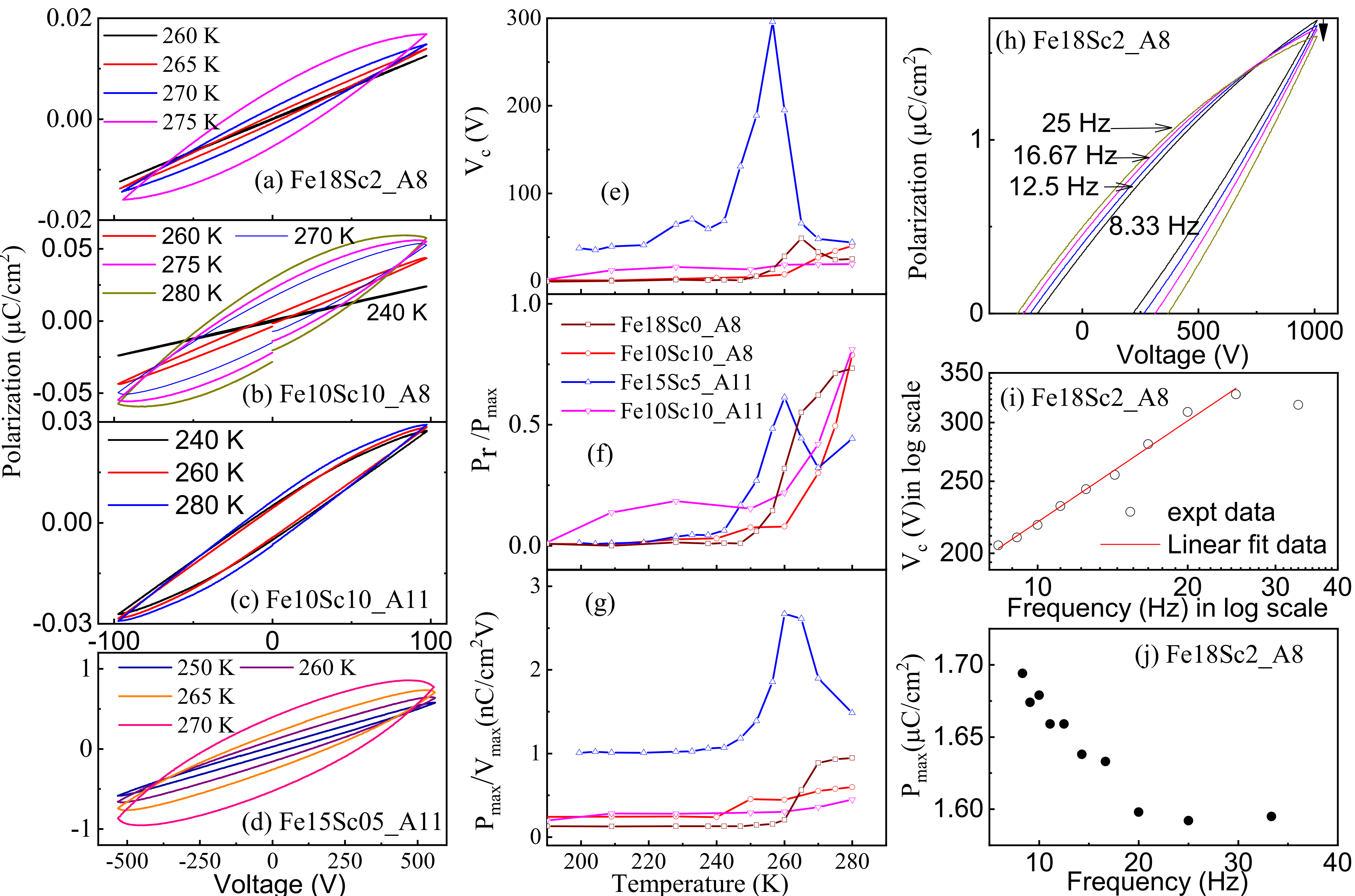

Fig. 7 P-V loops of the samples at selected temperatures (a-e), temperature dependence of the loop parameters (f-h), P-V loops at selected frequencies (h), frequency scaling of coercivity (i), frequency dependence of $P_{max}$ (j).

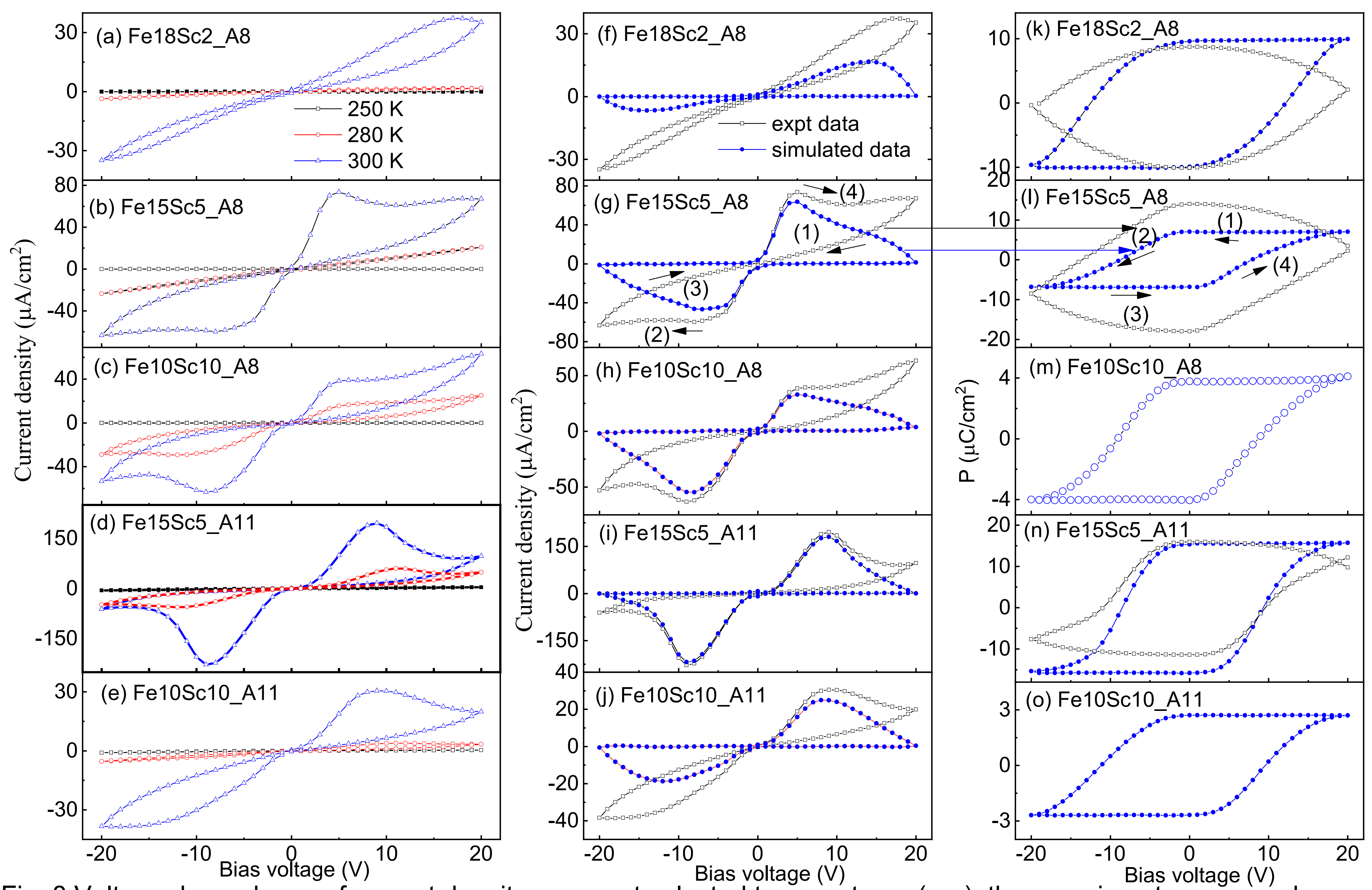

Fig. 8 Voltage dependence of current density curves at selected temperatures (a-e), the experiment curves and simulated curves at 300 K (f-j) and derived polarization curves (k-o). The arrows in (g) guides the direction of voltage.

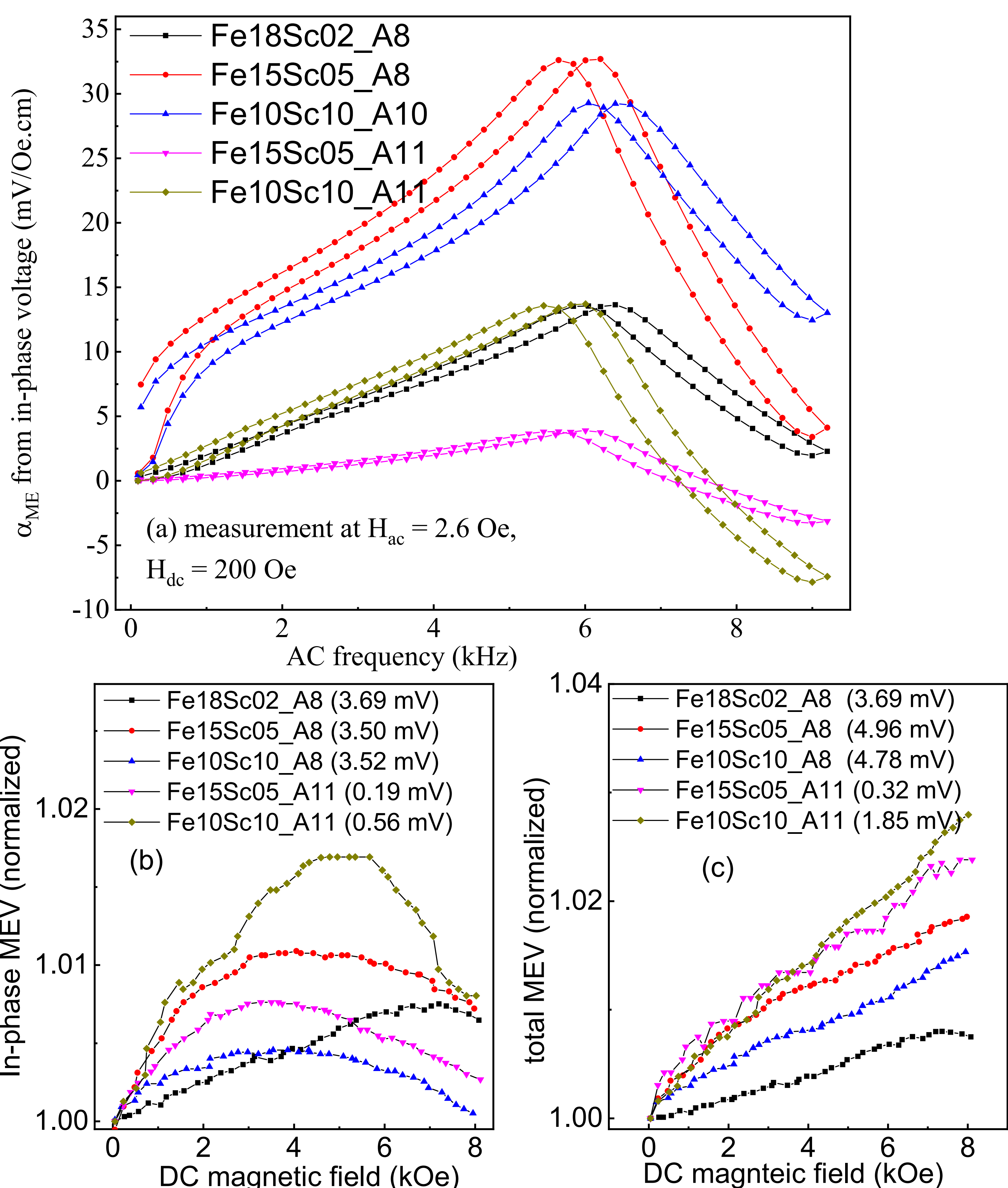

Fig. 9 Frequency dependence of the ME coupling coefficient using in-phase ME voltage and measured at dc field of 200 Oe (a), dc mganetic field dependence of the in-phase ME voltage (b) and total ME voltage (c) measured at ac field of amplitude 10 Oe and frequency 1997 Hz. The curves in (b-c) were normalized by their values at dc field of 50 Oe and showin within the brackets in the figures.

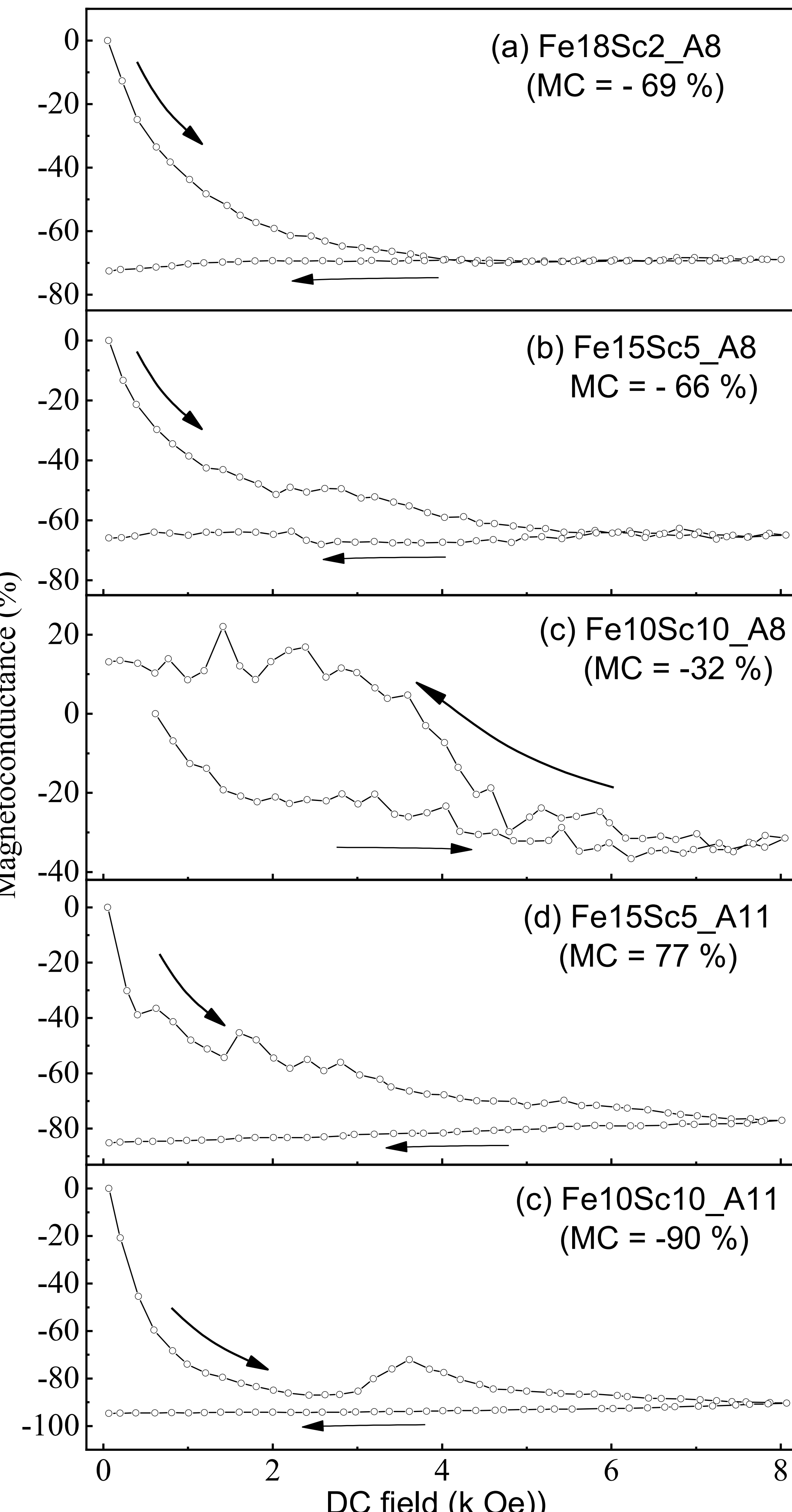

Fig. 10 Maagnetoconductance (MC) of the samples for magnetic field sweeping from 0 Oe to 8 kOe and back to 0 Oe.

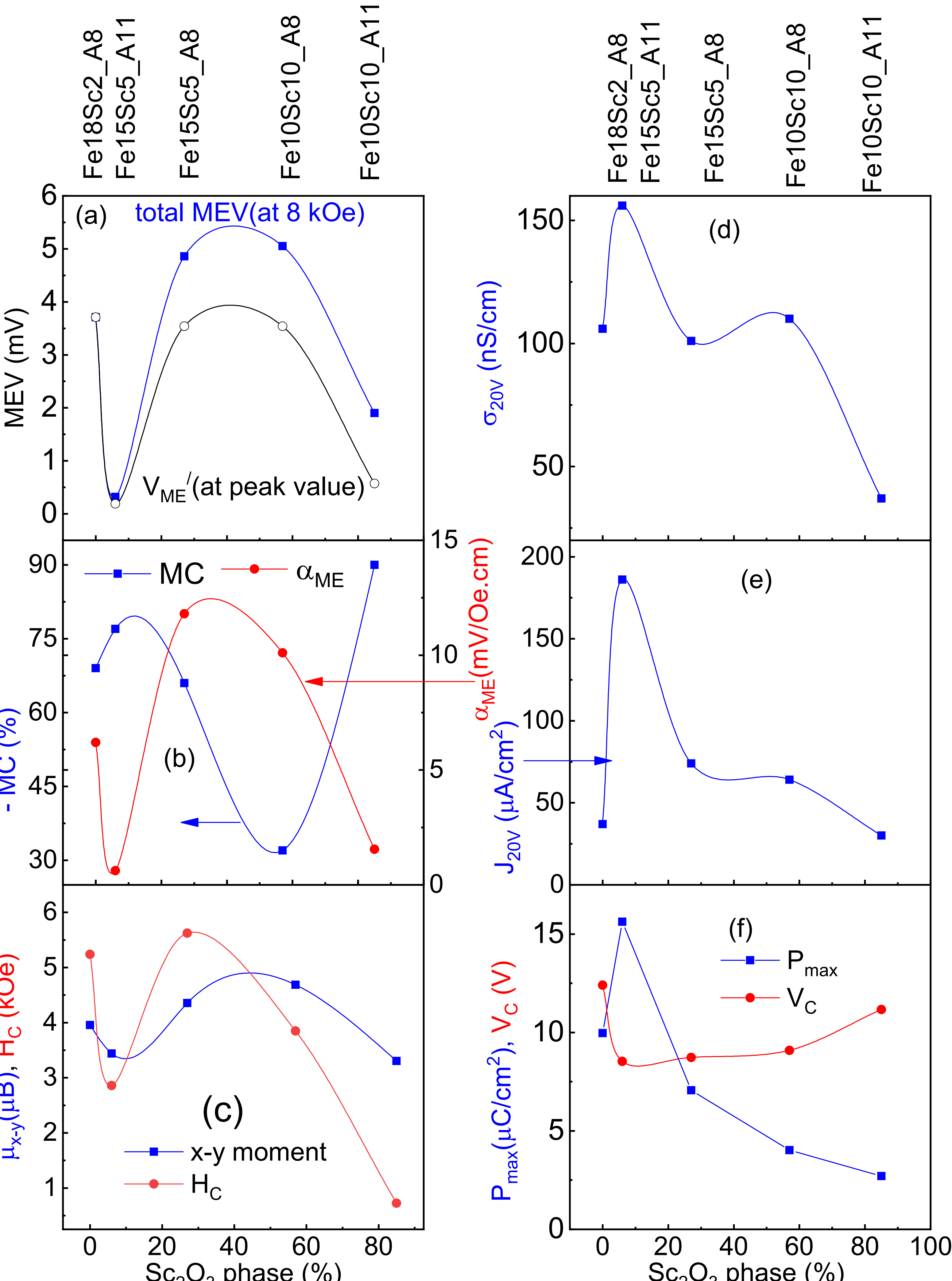

Fig. 11 Variation of the magnetoelectric, magnetic and electrical parameters at room temperature with $Sc_2O_3$ phase fraction in the samples.

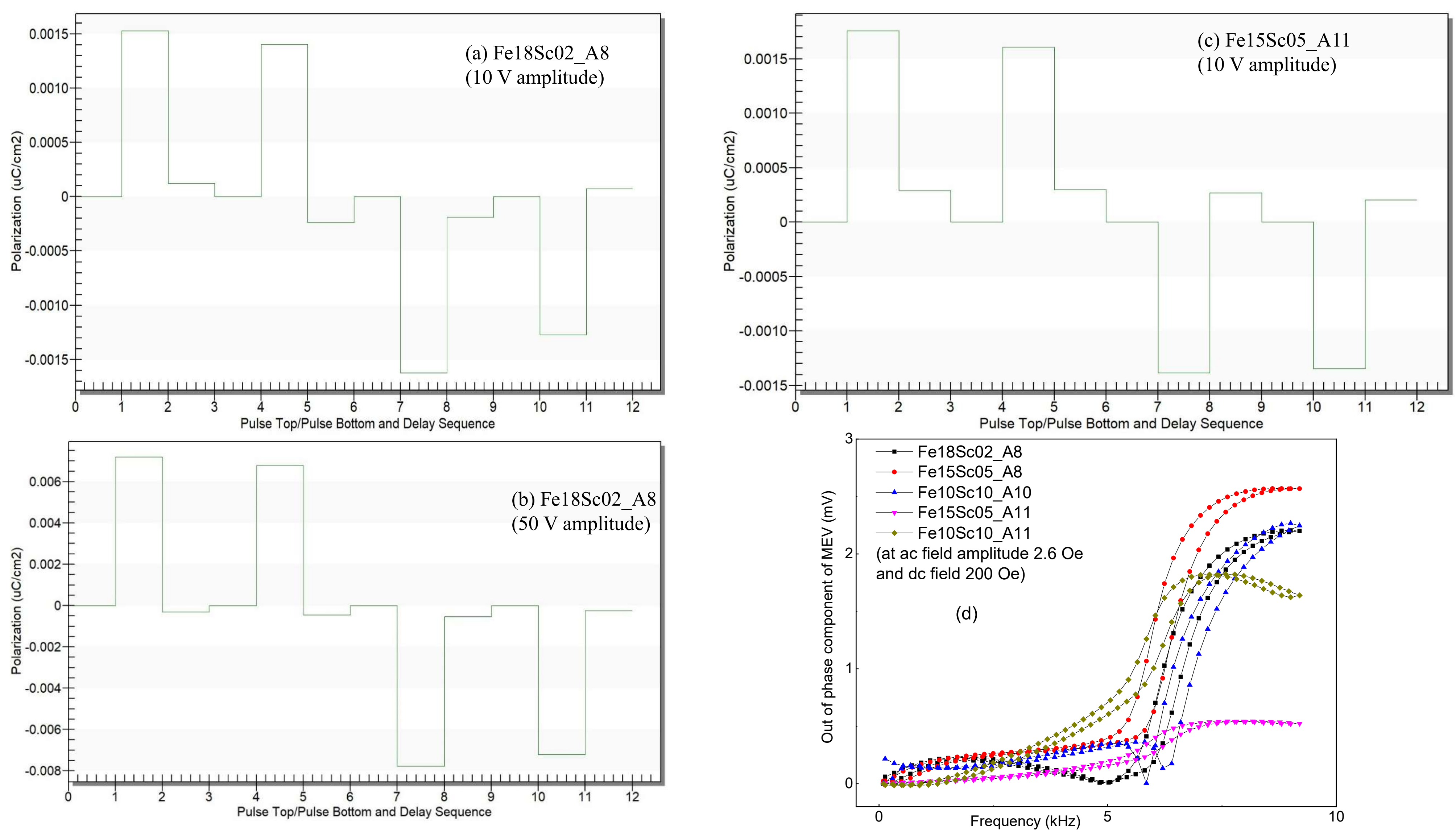

Fig. S1 PUND measurement data (a-c) and frequency scan of the out of phase ME signal (d) measured for different samples.

Table 1. The Fe–O bond-parameters (length ($d_{Fe-O}$), and angles (∠Fe–O–Fe and ∠O–Fe–O) from refined Neutron diffraction pattern of all samples.

| Samples | Temperature (K) | Bond length (Å) | | | Bond angle (degree) | | | | |
|---|---|---|---|---|---|---|---|---|---|
| | | Fe–O (×3) | Fe–O (×3) | <Fe–O> | ∠Fe–O–Fe | ∠Fe–O–Fe | ∠O–Fe–O | ∠O–Fe–O | ∠O–Fe–O |
| | 30 | 2.1125(13) | 1.9495(8) | 2.0310(4) | 94.05(5) | 131.63(7) | 90.59(6) | 102.66(4) | 161.99(11) |
| | 50 | 2.1151(14) | 1.9495(8) | 2.0323(5) | 94.05(5) | 131.59(7) | 90.54(5) | 102.68(4) | 162.00(6) |
| | 100 | 2.1158(14) | 1.9478(8) | 2.0318(4) | 94.04(5) | 131.57(7) | 90.52(5) | 102.70(4) | 161.99(11) |
| M2Sc02_A8 | 150 | 2.1144(14) | 1.9493(8) | 2.0319(4) | 94.03(5) | 131.61(7) | 90.54(5) | 102.64(4) | 162.07(11) |
| | 200 | 2.1149(17) | 1.9488(10) | 2.0318(5) | 94.03(7) | 131.59(8) | 90.53(6) | 102.65(4) | 162.06(15) |
| | 250 | 2.1163(14) | 1.9493(8) | 2.0328(5) | 94.04(5) | 131.58(7) | 90.51(5) | 102.66(4) | 162.07(11) |
| | 300 | 2.1168(13) | 1.9527(8) | 2.0348(4) | 94.06(5) | 131.62(7) | 90.56(5) | 102.68(4) | 161.99(11) |
| | 30 | 2.169(5) | 1.910(3) | 2.0300(15) | 94.6(3) | 130.17(16) | 90.9(4) | 102.60(14) | 161.9(4) |
| | 50 | 2.164(5) | 1.911(3) | 2.0295(15) | 94.6(3) | 130.31(16) | 90.7(4) | 102.54(14) | 162.1(4) |
| | 100 | 2.151(5) | 1.917(3) | 2.0280(12) | 94.5(3) | 130.59(16) | 90.7(4) | 102.56(13) | 162.0(4) |
| M2Sc05_A8 | 150 | 2.143(5) | 1.921(3) | 2.0269(15) | 94.4(3) | 130.77(16) | 90.0(4) | 102.50(2) | 162.0(4) |
| | 200 | 2.146(5) | 1.920(3) | 2.0274(15) | 94.4(3) | 130.72(16) | 90.7(4) | 102.57(13) | 162.0(4) |
| | 250 | 2.145(5) | 1.921(3) | 2.0274(15) | 94.4(3) | 130.75(16) | 90.7(4) | 102.5(2) | 162.0(4) |
| | 300 | 2.146(5) | 1.923(3) | 2.0284(15) | 94.4(3) | 130.73(16) | 90.0(4) | 102.7(2) | 162.5(4) |
| | 30 | 2.1134(15) | 1.9441(9) | 2.0288(5) | 94.07(5) | 131.54(8) | 90.52(5) | 102.78(4) | 161.84(12) |
| | 50 | 2.1136(16) | 1.9444(9) | 2.0290(5) | 94.07(6) | 131.55(9) | 90.54(6) | 102.79(4) | 161.82(12) |
| | 100 | 2.1031(16) | 1.9518(9) | 2.0274(5) | 94.05(6) | 131.79(9) | 90.76(6) | 102.59(4) | 161.98(12) |
| M2Sc10_A8 | 150 | 2.0994(16) | 1.9539(10) | 2.0266(5) | 94.00(6) | 131.88(10) | 90.81(6) | 102.42(4) | 162.22(12) |
| | 200 | 2.1088(16) | 1.9475(9) | 2.0281(5) | 94.03(6) | 131.66(9) | 90.63(6) | 102.64(4) | 161.98(12) |
| | 250 | 2.0918(19) | 1.9671(12) | 2.0294(6) | 94.06(9) | 132.30(12) | 91.13(7) | 101.86(5) | 162.86(12) |
| | 300 | 2.1120(2) | 1.9486(12) | 2.0304(6) | 94.02(8) | 131.64(12) | 90.63(9) | 101.48(7) | 161.94(15) |
| | 10 | 2.1185(5) | 1.9715(3) | 2.045(3) | 94.26(2) | 131.74(5) | 90.56(2) | 102.68(8) | 162.13(11) |
| | 50 | 2.1191(3) | 1.9720(7) | 2.0456(6) | 94.26(3) | 131.74(5) | 90.56(6) | 102.68(1) | 162.13(12) |
| | 100 | 2.1191(4) | 1.9720(4) | 2.045(6) | 94.25(7) | 131.75(4) | 90.56(5) | 102.67(9) | 162.12(9) |
| M2Sc05_A11 | 150 | 2.1190(3) | 1.9719(5) | 2.0455(7) | 94.25(4) | 131.74(5) | 90.56(4) | 102.67(9) | 162.12(9) |
| | 200 | 2.1195(5) | 1.9723(8) | 2.0460(4) | 94.25(5) | 131.74(6) | 90.56(8) | 102.67(6) | 162.12(8) |
| | 250 | 2.1195(4) | 1.9724(1) | 2.0460(8) | 94.25(6) | 131.74(6) | 90.56(6) | 102.67(8) | 162.12(9) |
| | 300 | 2.1197(8) | 1.9725(7) | 2.0461(3) | 94.25(8) | 131.74(5) | 90.56(4) | 102.67(9) | 162.12(9) |
| | 10 | 2.311(6) | 1.859(5) | 2.0853(19) | 91.3(2) | 127.0(5) | 88.7(2) | 100.3(2) | 168.4(2) |
| | 50 | 2.322(7) | 1.854(5) | 2.0879(25) | 91.5(3) | 126.9(5) | 88.5(2) | 100.9(2) | 167.7(3) |
| | 100 | 2.321(8) | 1.854(6) | 2.0878(24) | 91.4(3) | 126.8(6) | 88.6(3) | 100.8(3) | 167.9(3) |
| M2Sc10_A11 | 150 | 2.310(8) | 1.863(6) | 2.0862(28) | 91.3(3) | 127.1(6) | 88.7(3) | 100.2(3) | 168.6(3) |
| | 200 | 2.317(8) | 1.857(6) | 2.0873(24) | 91.4(3) | 126.9(6) | 88.6(3) | 100.6(3) | 168.1(3) |
| | 250 | 2.313(9) | 1.860(6) | 2.0867(31) | 91.4(3) | 127.0(7) | 88.6(3) | 100.5(3) | 168.2(3) |
| | 300 | 2.312(7) | 1.860(5) | 2.0858(22) | 91.4(3) | 127.0(5) | 88.6(2) | 100.4(2) | 168.4(3) |

Table 2. Magnetic parameters for the samples from M(H) loop measurements at 5 K/10 K and 300 K/350 K.

| Sample | | $H_{c1}$ | $H_{c2}$ | $H_c$ | $H_o$ | $M_{R1}$ | $M_{R2}$ | $M_R$ | $M_o$ | M | Shift of parameters |
|---|---|---|---|---|---|---|---|---|---|---|---|
| unit | | kOe | | | | (emu/g) | | | | | |
| M2Sc02_A8 | ZFC (10 K) | -4.018 | 3.067 | 3.542 | -0.475 | +0.100 | -0.067 | 0.084 | +0.016 | 1.344 (50 kOe) | |
| | ZFC (300 K) | -5.777 | 5.112 | 5.444 | -0.332 | 0.196 | -0.161 | 0.178 | +0.017 | 0.692 (20 kOe) | |
| M2Sc05_A8 | ZFC (10 K) | -0.559 | +0.397 | 0.478 | - 0.081 | +0.0165 | -0.0112 | 0.014 | +0.003 | 1.21 (50 kOe) | |
| | ZFC (350 K) | -5.372 | +5.879 | 5.625 | +0.253 | +0.113 | -0.113 | 0.113 | 0.000 | 0.818 (50 kOe) | |
| M2Sc10_A8 | ZFC (5 K) | -0.196 | 0.201 | 0.1985 | +0.0025 | +0.028 | -0.030 | 0.029 | -0.001 | 6.380 (70 kOe) | $H_{exb}$ = - 57 Oe<br>$\Delta H_c$ = 9 Oe<br>$M_{exb}$ = + 8.4 memu/g<br>$\Delta M_R$ = 2 memu/g |
| | FC (5 K) | -0.2616 | +0.1516 | 0.2066 | -0.055 | +0.0383 | -0.0234 | 0.031 | +0.0074 | 6.380 (70 kOe) | |
| | ZFC/FC (300 K) | -3.68 | +4.02 | 3.85 | +0.17 | +0.197 | -0.209 | 0.203 | -0.006 | 2.233 (70 kOe) | |
| M2Sc05_A11 | ZFC/FC (5 K) | -3.217 | +3.201 | 3.210 | 0.00 | +0.279 | -0.271 | 0.274 | 0.00 | 3.082 | $H_{exb} = \Delta H_c = 0$<br>$M_{exb} = \Delta M_R = 0$ |
| | ZFC (300 K) | -2.812 | +2.908 | 2.86 | +0.048 | +0.237 | -0.244 | 0.240 | -0.003 | 2.444 (70 kOe) | |
| M2Sc10_A11 | ZFC (5 K) | -0.718 | +0.856 | 0.787 | +0.069 | +0.105 | -0.115 | 0.11 | -0.005 | 8.273 (70 kOe) | $H_{exb}$ = - 354 Oe<br>$\Delta H_c$ = 153 Oe<br>$M_{exb}$ = +53 memu/g<br>$\Delta M_R$ = 25 memu/g |
| | FC (5 K) | -1.225 | +0.655 | 0.94 | -0.285 | +0.184 | -0.087 | 0.135 | +0.048 | 8.273 (70 kOe) | |
| | ZFC/FC (300 K) | -0.725 | +0.725 | 0.725 | 0 | +0.047 | -0.047 | 0.047 | 0 | 3.980 | |

Table 3. Off-plane canting angle (θ) of the Fe moments at 300 K and low temperature (5 K) for the Sc doped $Fe_2O_3$ system.

| | 5 K | | | 300 K | | |
|---|---|---|---|---|---|---|
| Sample | M(emu/g) | $\mu_x$ ($\mu_B$) | θ (deg) | M (emu/g) | $\mu_x$ ($\mu_B$) | θ (deg) |
| M2Sc02_A8 | 1.291 | 2.773 | 0.48 | 1.356 | 3.956 | 0.31 |
| M2Sc05_A8 | 0.932 | 1.559 | 0.63 | 0.698 | 4.356 | 0.17 |
| M2Sc10_A8 | 2.80 | 1.592 | 2.69 | 0.905 | 4.687 | 0.29 |
| M2Sc05_A11 | 1.87 | 3.456 | 0.57 | 1.44 | 3.44 | 0.44 |
| M2Sc10_A11 | 3.47 | 3.392 | 1.56 | 0.177 | 3.305 | 0.08 |

Note: The M at 50 kOe has been calculated from the M(H) curves and the $\mu_x$ ($\mu_B$) at 5 K has been estimated from the extrapolation of the $\mu_x$ (T) curves, determined from Rietveld refinement of the Neutron diffraction patterns of the samples.

Table S1. Wyckoff positions, occupancy and chemical composition of the atoms for all the samples.

| Sample code | T (K) | Space group | Atoms (site) | Wyckoff positions | | | B | Normalized occupancy | Chemical compositions |
|---|---|---|---|---|---|---|---|---|---|
| | | | | x | y | z | | | |
| Fe18Sc02_A8 | 30 | $R\overline{3}c$ | Fe (12c) | 0.00000 | 0.00000 | 0.35536(8) | 0.388(07) | 0.303(1) | |
| | | | Sc (12c) | 0.00000 | 0.00000 | 0.35536(8) | 0.526(09) | 0.036(3) | $Fe_{1.81}Sc_{0.21}O_3$ |
| | | | O (18e) | 0.3047(32) | 0.00000 | 0.25000 | 0.603(02) | 0.500 | |
| | 300 | $R\overline{3}c$ | Fe (12c) | 0.00000 | 0.00000 | 0.35539(7) | 1.871(1) | 0.320(1) | |
| | | | Sc (12c) | 0.00000 | 0.00000 | 0.35539(7) | 1.733(8) | 0.030(1) | $Fe_{1.92}Sc_{0.18}O_3$ |
| | | | O (18e) | 0.30481(33) | 0.00000 | 0.25000 | 1.536(3) | 0.500 | |
| Fe15Sc5_A8 | 30 | $R\overline{3}c$ | Fe (12c) | 0.00000 | 0.00000 | 0.35531 | 0.063(46) | 0.250 | |
| | | | Sc (12c) | 0.00000 | 0.00000 | 0.35531 | 0.201(46) | 0.083 | $Fe_{1.5}Sc_{0.49}O_3$ |
| | | | O (18e) | 0.30679(50) | 0.00000 | 0.25591(46) | 0.126(59) | 0.500 | |
| | | $Ia\overline{3}$ | Sc (8b) | 0.25000 | 0.25000 | 0.25000 | 0.635(10) | 0.167 | |
| | | | Sc (24d) | 0.46830(79) | 0.00000 | 0.25000 | 0.635(24) | 0.500 | |
| | | | O (48e) | 0.39373(50) | 0.14952(54) | 0.37812(46) | 0.399(59) | 1.000 | |
| | 300 | $R\overline{3}c$ | Fe (12c) | 0.00000 | 0.00000 | 0.35531 | 0.010(47) | 0.250 | |
| | | | Sc (12c) | 0.00000 | 0.00000 | 0.35531 | 0.128(47) | 0.083 | $Fe_{1.5}Sc_{0.49}O_3$ |
| | | | O (18e) | 0.30596(44) | 0.00000 | 0.25399(50) | 0.261(54) | 0.500 | |
| | | $Ia\overline{3}$ | Sc (8b) | 0.25000 | 0.25000 | 0.25000 | 0.486(93) | 0.167 | |
| | | | Sc (24d) | 0.46776(41) | 0.00000 | 0.25000 | 0.486(93) | 0.500 | |
| | | | O (48e) | 0.39290(44) | 0.15023(20) | 0.37620(50) | 0.264(54) | 1.000 | |
| Fe10Sc10_A8 | 30 | $R\overline{3}c$ | Fe (12c) | 0.00000 | 0.00000 | 0.35556(11) | 0.090(39) | 0.170(4) | |
| | | | Sc (12c) | 0.00000 | 0.00000 | 0.35556(11) | 0.228(39) | 0.166(7) | $Fe_{1.02}Sc_{0.99}O_3$ |
| | | | O (18e) | 0.30513(27) | 0.00000 | 0.25000 | 0.119(41) | 0.500 | |
| | | $Ia\overline{3}$ | Sc (8b) | 0.25000 | 0.25000 | 0.25000 | 0.988(65) | 0.182(2) | |
| | | | Sc (24d) | 0.46432(18) | 0.00000 | 0.25000 | 0.988(65) | 0.500(8) | |
| | | | O (48e) | 0.39310(27) | 0.15257(42) | 0.38197(36) | 0.633(70) | 1.000 | |
| | 300 | $R\overline{3}c$ | Fe (12c) | 0.00000 | 0.00000 | 0.35553(16) | 0.150(72) | 0.169(3) | |
| | | | Sc (12c) | 0.00000 | 0.00000 | 0.35553(16) | 0.012(72) | 0.167(8) | $Fe_{1.01}Sc_{1.00}O_3$ |
| | | | O (18e) | 0.30526(41) | 0.00000 | 0.25000 | 0.281(79) | 0.500 | |
| | | $Ia\overline{3}$ | Sc (8b) | 0.25000 | 0.25000 | 0.25000 | 0.081(59) | 0.185(4) | |
| | | | Sc (24d) | 0.46519(22) | 0.00000 | 0.25000 | 0.081(59) | 0.501(5) | |
| | | | O (48e) | 0.39323(41) | 0.15168(55) | 0.37656(51) | 0.157(75) | 1.000 | |
| Fe15Sc5_A11 | 10 | $R\overline{3}c$ | Fe (12c) | 0.00000 | 0.00000 | 0.35493 | 0.126 | 0.250 | |
| | | | Sc (12c) | 0.00000 | 0.00000 | 0.35493 | 0.012 | 0.083 | $Fe_{1.5}Sc_{0.49}O_3$ |
| | | | O (18e) | 0.30357 | 0.00000 | 0.25000 | 0.003 | 0.500 | |
| | | $Ia\overline{3}$ | Sc (8b) | 0.25000 | 0.25000 | 0.25000 | 0.483 | 0.167 | |
| | | | Sc (24d) | 0.46778 | 0.00000 | 0.25000 | 0.483 | 0.500 | |
| | | | O (48e) | 0.39289 | 0.15024 | 0.37621 | 0.264 | 1.000 | |
| | 300 | $R\overline{3}c$ | Fe (12c) | 0.00000 | 0.00000 | 0.35493 | 0.126 | 0.250 | |
| | | | Sc (12c) | 0.00000 | 0.00000 | 0.35493 | 0.012 | 0.083 | $Fe_{1.5}Sc_{0.49}O_3$ |
| | | | O (18e) | 0.30357 | 0.00000 | 0.25000 | 0.003 | 0.500 | |
| | | $Ia\overline{3}$ | Sc (8b) | 0.25000 | 0.25000 | 0.25000 | 0.483 | 0.167 | |
| | | | Sc (24d) | 0.46778 | 0.00000 | 0.25000 | 0.483 | 0.500 | |
| | | | O (48e) | 0.39289 | 0.15024 | 0.37621 | 0.264 | 1.000 | |
| Fe10Sc10_A11 | 10 | $R\overline{3}c$ | Fe (12c) | 0.00000 | 0.00000 | 0.35595(85) | 0.431 | 0.169(3) | |
| | | | Sc (12c) | 0.00000 | 0.00000 | 0.35595(85) | 0.293 | 0.168(3) | $Fe_{1.01}Sc_{1.00}O_3$ |
| | | | O (18e) | 0.35457(32) | 0.00000 | 0.25000 | 0.059 | 0.500 | |
| | | $Ia\overline{3}$ | Sc (8b) | 0.25000 | 0.25000 | 0.25000 | 0.126 | 0.182(2) | |
| | | | Sc (24d) | 0.46493(13) | 0.00000 | 0.25000 | 0.126 | 0.503(5) | |
| | | | O (48e) | 0.39250(32) | 0.15485(33) | 0.38016(32) | 0.323 | 1.000 | |
| | 300 | $R\overline{3}c$ | Fe (12c) | 0.00000 | 0.00000 | 0.35613(87) | 0.431 | 0.171(2) | |
| | | | Sc (12c) | 0.00000 | 0.00000 | 0.35613(87) | 0.293 | 0.160(5) | $Fe_{1.02}Sc_{0.96}O_3$ |
| | | | O (18e) | 0.35439(33) | 0.00000 | 0.25000 | 0.059 | 0.500 | |
| | | $Ia\overline{3}$ | Sc (8b) | 0.25000 | 0.25000 | 0.25000 | 0.126 | 0.181(4) | |
| | | | Sc (24d) | 0.46523(13) | 0.00000 | 0.25000 | 0.126 | 0.503(9) | |
| | | | O (48e) | 0.39232(33) | 0.15541(34) | 0.37987(33) | 0.323 | 1.000 | |